\documentclass[preprintnumbers,superscriptaddress,floatfix,showpacs,prd]{revtex4}

\usepackage{longtable}
\usepackage{amsmath,amssymb,bm}
\usepackage{graphicx}
\usepackage{hyperref}
\usepackage{amsfonts}
\usepackage{latexsym}
\usepackage{amssymb}
\usepackage[caption=false]{subfig}
\usepackage{color}

\begin{document}	
\title{Causality and Stability Conditions of a Conformal Charged Fluid}
\author{Farid Taghinavaz}
\affiliation{School of Particles and Accelerators, Institute for Research in Fundamental Sciences (IPM), P.O. Box 19395-5531, Tehran, Iran}	
\email{ftaghinavaz@ipm.ir}
		\begin{abstract}
			In this paper, I  study the conditions imposed on a normal charged fluid so that the causality and stability criteria hold for this fluid. I adopt the newly developed General Frame (GF) notion in the relativistic hydrodynamics framework which states that hydrodynamic frames have to be fixed after applying the stability and causality conditions. To my purpose, I take a charged conformal matter in the flat and $3+1$ dimension to analyze better these conditions. The causality condition is applied by looking to the asymptotic velocity of sound hydro modes at the large wave number limit and stability conditions are imposed by looking to the imaginary parts of hydro modes as well as the Routh-Hurwitz criteria.  By fixing some of the transports, the suitable spaces for other ones are derived. I have observed that in a dense medium with finite $U(1)$ charged chemical potential $\mu_0$, negative values for transports appear and the second law of thermodynamics has not ruled out the existence of such values. Sign of scalar transports are not limited by any constraints and just a combination of vector transports is limited by the second law of thermodynamic.  Also numerically it is proved that the most favorable region for transports $\tilde{\gamma}_{1, 2}$, coefficients of the dissipative terms of the current, is of negative values. 
		\end{abstract}
				\maketitle
	\section{Introduction}
	Relativistic Hydrodynamics (RH) is the most powerful tool to describe the in and out of equilibrium properties of hot and dense QCD matter. It explains the dynamics of quark matter in terms of some local effective Degrees of Freedom (DoF) rather than infinite microscopic DoF. Examination of RH has been intensified since the previous decade  which experiments confirm that observables of heavy ion collision are in good agreement with the predictions of RH \cite{Romatschke:2017ejr, Florkowski:2017olj, Jeon:2015dfa, Kovtun:2012rj}. Another framework to study the features of Quark-Gluon Plasma (QGP) is the Relativistic Kinetic Theory (RKT). There are some differences between these two frameworks. The First one is that the RH can be studied for either the weakly and strongly interacting field theories, while the RKT is only applicable for weakly coupled field theories which quasi-particles have good definition. The second one is that the RKT is insensitive to the momenta regime, while for the RH the slowly varying assumption of macroscopic fields has a major priority and due to this, the RH is based upon the gradient expansion of macroscopic fields which favors the low momenta region.\\
	In the last few years because of some experimental and analytical challenges, the studies of RH has changed its path. Here, I am going to describe two of these challenges and try to put my problem in one of these mainstreams.	Traditionally, it was believed that two conditions are mandatory for the RH. The first one is having  a local and stable thermal equilibrium and the second one is the validity of  gradient expansion as a consequence of the slowly varying assumption. In order to  achieve to these axioms, we have to deal with large number of colliding particles. Otherwise, the concept of equilibrium and slow variation do not make sense. But the recent observations in RHIC and LHC has suspected us about these two axioms. These observations are about the collective behaviors in small system collisions in which the local thermal equilibrium state and the smooth variation of macroscopic fields stop to reach \cite{Aad:2012gla, CMS:2012qk, Abelev:2012ola, Khachatryan:2016txc}. After these observations, the theoretical works has pushed to study  the late time behavior of QCD matter dynamics. There is large literature in this field which states that an attractor solution appears in the hydrodynamics calculations regardless of any initial conditions,  \cite{Heller:2020anv, Heller:2015dha, Heller:2018qvh, Heller:2016rtz, Heller:2013fn, Shokri:2020cxa, McNelis:2020jrn, Denicol:2018pak, Blaizot:2017ucy, Strickland:2017kux, Aniceto:2015mto, Basar:2015ava} and references therein. The meaning of this word is that RH can be applied to any high energetic collisions of particles -regardless of its size - which its equations of motion have an attractor blind to the initial conditions and the RH has its physical meaning at enough late times. In another word, RH arises when the initial non-hydrodynamics modes are damped. This is one way in the recent works of RH.\\
    Another stream in the RH is to work with the everlasting problem, namely the stability and causality issues. Historically, it has been proved that first order RH  suffers from the growing amplitudes of fluctuations in time (the stability problem) and the superluminal propagation speed of fluctuations (the causality problem) \cite{Hiscock:1985zz, Hiscock:1987zz, Israel:1976tn, Hiscock:1983zz}. This pathology has been cured by introducing the second order terms in the entropy current vector. Such a view, namely the inclusion of phenomenological second order terms into the first order dissipative calculations is known as the Muller-Israel-Stewart (MIS) framework. I have to emphasize that the MIS approach does not guarantee the stability and causality of hydro modes, per se. But rather, for the MIS theory to be a stable and causal formalism, it should satisfy some condition. For example $\eta$, the shear transport coefficient and $\tau$, the shear relaxation time are no longer independent parameters, but instead satisfy ${\eta \over \tau_\pi T s}\leq {1\over 2}$ \cite{Pu:2009fj}. Also the local velocity of fluid's parcel and dimension of space-time influence these conditions \cite{Denicol:2008ha}.  However, recently it appears a trend in this channel which does not need to include higher order gradient terms in the entropy current \cite{Kovtun:2019hdm, Bemfica:2019knx, Bemfica:2017wps}. These works have focused on the notion of General Frame (GF) and definition of new transport coefficients ahead of the gradient terms. In the natural process of the RH, when dissipative terms enter into the calculations, the concept of "frame" arises. This is because in  out of equilibrium cases, the thermodynamic fields such as temperature and chemical potential lose their meanings and we can not define the unique value for them \cite{Kovtun:2012rj}. We can vary locally the thermodynamical (thermo) fields without harming the RH equations. In some sense, it is equivalent to the gauge freedom in the QFT. People usually have used  this freedom to fix the frame, i.e. the Landau or Eckart frame, and then proceed to do the calculations. This is the old approach to the dissipative RH and it results to the unstable and acausal modes. In the new fashion, I mean in the GF approach, the  frame freedom is respected and we do not try to fix them prior to any calculations. We have to first perform the RH computations and since then decide which frame is physical and good for our purpose. For instance, the stability and causality can be studied in this way. We utilize  the notion of GF to fix the hydrodynamical frames after computing the hydrodynamical (hydro) modes. This job has shrunk the space of transport coefficients and reduced them to those satisfy some special conditions.\\
    We can compare the MIS and GF approach in two distinct ways. The MIS lacks  fundamental bases and it is only a phenomenological approach, but the GF frame is based upon the good deal with frame notion which is a physical concept. That is why the GF framework does not posses the unstable and acausal modes even by not including the artificial terms. Another difference between these two approaches is that the MIS enters some variables into the RH with an extra relaxation type equation, while in the GF approach there is no any DoF besides the temperature, chemical potential and fluids velocity. Therefore, in this manner, it seems that GF emerges from a renormalizable field theory while the MIS does not originate from a renormalizable theory.\\
    So far the investigations in the GF approach are about the uncharged conformal fluid and try to limit the space of transports to the causal and stable regions. My motivation to do this work is  to generalize the preceding works to a QCD plasma with finite charge density, equivalently finite chemical potential "$\mu_0$". Throughout this paper, by the chemical potential I mean the $U(1)$ charged chemical potential and discard any other ones. I do the calculations for two circumstances. First is for a hyperdense fluid with $\mu_0 \gg T_0$ and the second for a fluid with finite $\mu_0$ and $T_0$. Both of these studies are done by the assumption of conformal symmetry imposed on the Equation of State (EoS) and other related quantities. Motivation to split as this is in the QCD phase diagram. In the QCD phase diagram, the QGP phase can be seen in two distinct areas: i) high temperature region without any charge $T_0 \gg \mu_0$  and ii) regime which has finite $\mu_0$ and $T_0$. The hyper dense medium of quark matter is believed to be as color-superconductor phase and it is not a strongly interacting plasma but studying the hydrodynamics of this medium is of great importance. I shall try to generalize the previous works to the hyper dense and finite $T_0$ and $\mu_0$ medium and obtain the physical conditions on the transports which causes a conformal charged fluid to be a stable and causal theory. For conformal charged matter the number of transports are very large, the nine parameters and knowing all of the suitable spaces is a subtle job. Therefore, I fix five of them and derive the conditions in terms of rest four. Various parameter sets are taken and physical spaces are derived for the $\tilde{\gamma}_2$ and $\tilde{\epsilon}_2$ transports. In the finite density medium case this work is repeated for two values of ${\mu_0 \over T_0}$. The main achievement of this paper is that sign of scalar transports are not fixed by using the second law of thermodynamics $\left(\partial_{\mu} S^\mu \geq 0\right)$ and just sign of combination of vector transports are fixed by it. Thus, scalar transports as well as the vectors might have negative values and second law does not rule out it anymore. Importance of this issue is that we usually expect the transports have to be non-negative values and so far the negative values for transports are not seen. Therefore, appearance of negative values for transports seems to be a new challenge for RH. Another achievement of the current paper is that the regions derived from stability conditions satisfies the causality condition. In another word, stability gives causality.\\
    The organization of this paper is as follows. At sec II.) I shall discuss about some preliminary stuffs of the RH, idea of frame in the dissipative hydrodynamics and the notion of GF. At sec III.) I will try to get some conditions on the transport parameters of a hyper dense fluid $\mu_0 \gg T_0$ by taking the conformal symmetry over the EoS and other quantities. In order to derive the conditions, stability, causality and second law arguments are used simultaneously. The stability criteria is applied by looking to the imaginary parts of hydro modes as well as the Routh-Hurwitz criteria and causality condition is derived by looking to the asymptotic velocity of sound modes in the large wave number limit. By fixing five of them the physical regions for other transports are derived and the space of $\tilde{\gamma}_2$ and $\tilde{\epsilon}_2$ transports is constrained. Various parameter sets are taken to realize these zones. At section IV), I will repeat the works done in the section III) for finite $T_0$ and $\mu_0$ medium and for two values of ${\mu_0 \over T_0}$. Eventually, I close the paper with discussion  about the conclusions and  I address some problems which can be done along this paper.\\
    Throughout the present paper I take the $\hbar = c = k_B = 1$ convention and the Minkowski metric is chosen to be as $g_{\mu \nu} = diag(-1, 1, 1, 1)$.     
	\section{Preliminaries}
	Late time behavior of a relativistic system can be described by using the conserved currents. According to the Noether theorem, these currents belong to either the space-time symmetries or internal symmetries. Energy momentum tensor $T^{\mu \nu}(x)$ as a conserved current, corresponds to the space-time symmetries, while another conserved currents such as $J^\mu(x)$ correspond to the symmetries of internal space. In the cases without any anomalies and in $d+1$ space-time dimension, the number of independent components for energy momentum tensor and currents are ${(d+1)(d+2)\over 2}$ and $d+1$, respectively. To derive the dynamical evolution of the fluid, the following conservation laws are used
		\begin{align}\label{eqsec11}
		\nabla_\mu T^{\mu \nu}&=0,\\\label{eqsec12}
		\nabla_\mu J^\mu &=0,
		\end{align}
	which are consequences of diff and gauge transformations \cite{Jensen:2012kj}. The latter equations impose $d+1$ constraints on $T^{\mu \nu}(x)$ and one constraint on the $J^\mu(x)$ components. Therefore, it seems impossible to solve these equations completely unless we assume some physical conditions. For relativistic systems in the macroscopic level the conserved currents have to be written in terms of some local effective DoF. Number of these DoF are chosen to be as same as the number of conservation laws ($d+2$). Thus, in order to be consistent, the conserved currents should be expressed in terms of these $d+2$ DoF. The way of representing the conserved currents in terms of DoF is known as constitutive relations. Generally, the energy momentum tensor and current density of a given fluid are written in terms of DoF as the following ones
	\begin{align}\label{eqsec13}
	T^{\mu \nu}(x) &= \mathcal{E}(x) u^\mu(x) u^\nu(x) + \mathcal{P}(x) \Delta^{\mu \nu}(x) + Q^\mu(x) u^\nu(x) + Q^\nu(x) u^\mu(x) + t^{\mu \nu}(x),\\\label{eqsec14}
	J^\mu(x) &= \mathcal{N}(x) u^\mu(x) + \mathcal{J}^\mu(x).
	\end{align}
	In this relation, $\mathcal{E}(x), \mathcal{P}(x)$ and  $\mathcal{N}(x)$ represent the local energy density, pressure  and a conserved number density, respectively. Furthermore, $u^\mu(x)$ stands for the local fluid velocity and $\Delta^{\mu \nu}(x) = g^{\mu \nu}(x)+ u^\mu (x) u^\nu (x)$ is an operator which  projects a given tensor onto the space perpendicular to the  $u^\mu(x)$.   $Q^\mu(x)$ is the local heat current and $t^{\mu \nu}(x)$ is the traceless symmetric part of the energy momentum tensor.  $\mathcal{J}^\mu(x)$ is an added term to the current density which appears in higher derivative corrections. The fields  $Q^\mu(x), \mathcal{J}^\mu(x)$ and $t^{\mu \nu}(x)$ have two properties. The first one is that they are all transverse to the $u^\mu(x)$ and this constraint reduces the number of independent components in each of them. The second one is that they vanish in the ideal limit  where all derivative corrections are muted. I have described before that the conserved currents and consequently all the hydro fields such as $\left(\mathcal{E}(x), \mathcal{P}(x), Q^\mu(x), t^{\mu \nu}(x), \mathcal{N}(x), \mathcal{J}^\mu(x)\right)$ have to be described in terms of d+2 DoF. I choose these DoF to be as the $\left(T(x), \mu(x), u^\mu(x)\right)$. The local equilibrium values of these DoF are called as the thermo fields. We use the convention $u^\mu u_\mu=-1$ and that is why the number of independent components for $u^\mu$ is $d$.\\
	Besides the constitutive relations, to work with the RH we have to use  another important property and that is the derivative expansion assumption. In this assumption, the above mentioned hydro fields are expressed in terms of slowly varying thermo fields and their derivatives.  Indeed, it is assumed that thermo fields have small fluctuations in the region where thermodynamic is applicable and in the expansion procedure, each term is small compared to its preceding one. The general forms of hydro fields are written as it follows\cite{Kovtun:2012rj}
			\begin{align}\label{eqsec15}
			\mathcal{E} &= \epsilon_0 + f_{\mathcal{E}}\left(\partial T, \partial \mu, \partial u\right) + \ldots, \quad \mathcal{P} = p_{id} + f_{\mathcal{P}}\left(\partial T, \partial \mu, \partial u\right) + \ldots ,\quad \mathcal{N} = n_{0} + f_{\mathcal{N}}\left(\partial T, \partial \mu, \partial u\right) + \ldots \\\label{eqsec18}
			\mathcal{Q}^\mu &= f_{\mathcal{Q}}\left(\partial T, \partial \mu, \partial u\right) + \ldots , \qquad \,\,\mathcal{J}^\mu = f_{\mathcal{J}}\left(\partial T, \partial \mu, \partial u\right) + \ldots , \\
			 t^{\mu \nu} &= f_{t}\left(\partial T, \partial \mu, \partial u\right) + \ldots.
			\end{align}
		 For the sake of brevity I omit the $x$ dependence in these fields. In the latter relation, the notation $"\ldots"$ stands for the higher order corrections which stems from the higher derivative terms. At zeroth order when we deal with the ideal fluid, the equilibrium values of hydro fields are written in terms of  the local values of thermo fields
	\begin{align}\label{eqsec19}
		\epsilon_0 = \mathcal{E}(T_0, \mu_0, u^\mu_0), \qquad p_{id} = \mathcal{P}(T_0, \mu_0, u^\mu_0), \qquad n_{0} = \mathcal{N}(T_0, \mu_0, u^\mu_0).
	\end{align}
	 The zero sub indices  refer to the local thermal values of each field.  The functions $f\left(\partial T, \partial \mu, \partial u\right)$ represent the first order corrections to the constitutive relations. By going to higher derivative terms in the hydrodynamics calculations, the idea of frame plays an important role. It works as it follows. The thermo fields have no unique definitions in the higher order corrections. It means that we can redefine them by adding new contributions
	 \begin{align}\label{eqsec110}
	 T \to T + \delta T, \qquad \mu \to \mu + \delta \mu, \qquad u^\mu \to u^\mu + \delta u^\mu,
	 \end{align}
	  in such a way that energy momentum tensor and current density remain unchanged. Therefore, there is no any preferred values for these thermo fields \cite{Kovtun:2012rj}.  Different redefinitions of thermo fields are usually called as the "hydro frames" and the freedom in choice of  specific value for thermo fields is often called as the "frame freedom". At this level, these redefinitions  resemble to the gauge freedom in QFT.  Naturally, the corrections $\left(\delta T, \delta \mu, \delta u^\mu\right)$ have to be written in terms of derivatives of thermo fields\cite{Kovtun:2019hdm}
	  	\begin{align}\label{eqsec111}
	  	\delta T &=  a_1 {u^\mu \partial_{\mu} T\over T} + a_2 \partial_{\mu} u^\mu + a_3 u^\mu \partial_{\mu} \left(\mu \over T\right) +  \ldots, \\\label{eqsec112}
	  	\delta \mu &=  c_1 {u^\mu \partial_{\mu} T\over T} + c_2 \partial_{\mu} u^\mu + c_3 u^\mu \partial_{\mu} \left(\mu \over T\right)+ \ldots, \\\label{eqsec113}
	  	\delta u^\mu &=  b_1 u^\nu \partial_{\nu} u^\mu + b_2 {\Delta^{\mu \nu} \partial_{\nu} T\over T} + b_3 \Delta^{\mu \nu} \partial_{\nu} \left(\mu \over T\right)+ \ldots.
	  	\end{align}
	   The coefficients $\left(a_i, b_i, c_i\right)$ are arbitrary real numbers and the concept of frame is referred to choose some specific values for these numbers. Also the functions $\left({u^\mu \partial_{\mu} T\over T}, \partial_{\mu} u^\mu, u^\mu \partial_{\mu} \left(\mu \over T\right)\right)$ and $\left(u^\nu \partial_{\nu} u^\mu, {\Delta^{\mu \nu} \partial_{\nu} T\over T}, \Delta^{\mu \nu} \partial_{\nu} \left(\mu \over T\right)\right)$ are independent scalar and vector bases and are useful to expand another quantities in term of them. For example, we can write the hydro fields as a function of these bases 
	  	\begin{align}\label{eqsec114}
	  	\mathcal{E} &= \epsilon_0 + \sum_{i =1}^{3} \epsilon_i s_i, \qquad \mathcal{P} = p_{id} + \sum_{i =1}^{3} \pi_i s_i, \qquad \mathcal{N} = n_{0} + \sum_{i =1}^{3} \nu_i s_i, \\\label{eqsec115}
	  	\mathcal{Q}^\mu &= \sum_{i =1}^{3} \theta_i v_i^\mu, \qquad \qquad  \mathcal{J}^\mu =\sum_{i =1}^{3} \gamma_i v_i^\mu, \qquad \qquad t^{\mu \nu} = - \eta \sigma^{\mu \nu}.
	  	\end{align}
	  In the latter relations, $s_i, v_i^\mu$  and $\sigma^{\mu \nu}$ are scalar, vector and tensor expressions built out of derivatives of thermo fields
	  	\begin{align}\label{eqsec116}
	  	s_1 &= {u^\mu \partial_\mu T \over T}, \qquad  s_2 = \partial \cdot u, \qquad s_3 = u^\mu \partial_\mu \left({\mu \over T}\right),\\\label{eqsec117}
	  	v_1^\mu &= u^\alpha \partial_\alpha u^\mu, \qquad v_2^\mu = {\Delta^{\mu \alpha} \partial_\alpha T\over T}, \qquad v_3^\mu = \Delta^{\mu \alpha} \partial_\alpha \left({\mu \over T}\right),\\\label{eqsec118}
	  	\sigma^{\mu \nu}& = \mathcal{P}^{\mu \nu \alpha \beta} \partial_\alpha u_\beta, \qquad  \mathcal{P}^{\mu \nu \alpha \beta} \equiv {1\over 2} \left(\Delta^{\mu \alpha} \Delta^{\nu \beta} + \Delta^{\mu \beta} \Delta^{\nu \alpha} - {2\over 3} \Delta^{\mu \nu} \Delta^{\alpha \beta}\right).
	  	\end{align}
	  The numbers $\left(\epsilon_{i}, \pi_i, \nu_{i}, \theta_{i}, \gamma_{i}, \eta \right)$ are called as the transport coefficients and until no condition is implied, they  are arbitrary numbers. We can study the impact of thermo field redefinition on these transport coefficient \cite{Kovtun:2019hdm}. Choosing the scalar and vector sets are not unique and one can adopt another sets by making a linear combination. In the usual use of RH, I mean before the GF notion, people often have benefited of frame freedom defined in the relation \eqref{eqsec110}, to fix the frame prior to any calculation. But the idea of GF tells us that we have to keep the transports undetermined and proceed the computations and after doing them we fix the transports according to our needs.\\
	   In the next section I have used of this frame freedom in order to choose the appropriate scalar and vector bases to derive the hydro modes. The Difference of my works with the paper \cite{Kovtun:2019hdm} is that I take into account the influences of $"s_3"$ and $"v_3^\mu"$ in the hydro modes and study the stability and causality conditions implied by adding these new bases. I shall do this by changing the bases in the relations \eqref{eqsec116} and \eqref{eqsec117}. \\
	  It is worthwhile to mention that the notion of frame even works in the thermodynamic states where there is no any effect of dissipation terms \cite{Jensen:2012jh, Abbasi:2017tea}. In the current paper I limit myself to study the impact of frames in a charged dissipative fluid and postpone the study of thermodynamics frames to the future works. \\	  
	 	Hereafter, I split the calculations into two parts. The Next section is devoted to the calculations of hydro modes for very dense systems ($\mu_0 \gg T_0$) and investigating the stability and causality conditions implied on this fluid. The section IV) belongs to  the same calculations for finite $\mu_0$ and $T_0$.
	 	\section{Dense fluids}
	 By dense medium, I mean a very cold and charged medium which has the condition $\mu_0 \gg T_0$ of equilibrium values. we have to be careful about the hydrodynamics of dense medium, since there is a great difference between the hydrodynamics equations for the hot and uncharged medium and the hydrodynamics in the cold and dense medium. In the former case, we could safely adopt the following sets of thermodynamics states  and fluctuations  
	 \begin{align}\label{eqsec21}
	 	& \mbox{Thermodynamic state},\qquad \mu_0=0, \,\, T_0\neq 0, \,\, u^\mu_0= \left(1, 0,0,0\right),\\\label{eqsec22}
	 	& \mbox{Fluctuations},\qquad \qquad \qquad  \delta \mu=0, \,\, \delta T\neq 0, \,\, \delta u^\mu= \left(0, \delta u_x, \delta u_y, \delta u_z\right).
	 \end{align}
	 Using of these choices do not enter any flaw in our calculations. we have four equations (conservation laws of energy-momentum) for four unknown variables $\left(\delta T, \delta u^x, \delta u^y, \delta u^z\right)$ and they are solved consistently. In the latter case (hydrodynamics of cold and dense medium), we could not set the following thermodynamics states and fluctuations 
	 \begin{align}\label{eqsec23}
	 & \mbox{Thermodynamic state},\qquad T_0=0, \,\, \mu_0\neq 0, \,\, u^\mu_0= \left(1, 0,0,0\right),\\\label{eqsec24}
	 & \mbox{Fluctuations},\qquad \qquad \qquad  \delta T=0, \,\, \delta \mu\neq 0, \,\, \delta u^\mu= \left(0, \delta u_x, \delta u_y, \delta u_z\right).
	 \end{align}
	 This is because we have five equations (conservation laws of energy-momentum and charge) for four unknown fluctuations $\left(\delta \mu, \delta u^x, \delta u^y, \delta u^z\right)$. Therefore,  the hydrodynamic equations of cold and dense medium instead of using the relations \eqref{eqsec23} and \eqref{eqsec24}, is started with the following sets 
	 \begin{align}\label{eqsec25}
	 & \mbox{Thermodynamic state},\qquad T_0=0, \,\, \mu_0\neq 0, \,\, u^\mu_0= \left(1, 0,0,0\right),\\\label{eqsec26}
	 & \mbox{Fluctuations},\qquad \qquad \qquad  \delta T \neq 0, \,\, \delta \mu\neq 0, \,\, \delta u^\mu= \left(0, \delta u_x, \delta u_y, \delta u_z\right).
	 \end{align}
	 This seems to be physical, since chemical potential is only a parameter which adjusts the energy scale to create a charged particle. So, in an uncharged medium there is no reason to have non vanishing  chemical potential fluctuation. However, temperature is the more fundamental parameter. we might think of a very cold medium $(T_0=0)$, but we are not able to turn off its fluctuation, since thermal fluctuations are related to the motion of particles and they even exist in the very cold medium. Therefore, the chosen sets in equations \eqref{eqsec25} and \eqref{eqsec26} are suitable for our problem. By using these choices, we are lead to the consistent equations which do not need further information.\\
	 It should be emphasized here that the choice of fluid's velocity $u^\mu_0= \left(1, 0,0,0\right)$ is compatible with the Bjorken symmetry. Indeed, one can show that the symmetry arguments constrain the form of fluid's velocity. Bjorken symmetry is a special symmetry pattern which fit well to the experimental observables at the initial stages of heavy ions collisions. It comprises of four symmetries including the boost, rotation along the $z$ direction and two translational symmetries ($\partial_x, \partial_y$). Only the three symmetries are independent of each other which consequently determine the velocity form.\\
	 In order to derive the stability and causality conditions, I try to obtain the hydro modes. By discussing over these modes, I are able to derive the conditions over the transport coefficients which makes the first order charged fluid to be a stable and causal theory. Before going into the details of calculations, I point out that the following calculations of hydro modes are done for fluid at its local rest frame. One can  do the same calculation for locally boosted fluids $u^\mu = {1\over \sqrt{1-v^2}}\left(1, v^i\right)$, only by a simple boost transformation between the solutions in local rest frame and boosted frame \cite{Kovtun:2019hdm}. However, having the non zero velocity might cause changing the quality of conditions \cite{Denicol:2008ha}. I shall show that by using the asymptotic causality condition, I obtain exact and correct results for parameter space of transport coefficients. \\
     Here, I mention briefly on hydro modes. To derive the hydro modes, I have to perform some steps. Firstly, I should set our thermodynamic states and fluctuations. I call the thermal fields as $\phi_a$ and its fluctuations as $\delta \phi_a$. As described before for dense fluids, I we select the following  collections
     \begin{equation}\label{eqsec27}
     	\phi_a + \delta \phi_a = \left(\delta T, \, \mu_0 + \delta \mu, \, u_0^\mu + \delta u^\mu\right).
     \end{equation}
     The fields $\phi_a$ as the thermal states are defined in the local rest frame of fluid. After setting these states, I have to perturb the constitutive relations up to the first order in fluctuations. It means that the shown thermo fields in the relation \eqref{eqsec27} have to be inserted in the constitutive relations of currents and energy momentum tensor and expand them up to first order in fluctuations. In the case of dense fluids, constitutive relations are like as the relations \eqref{eqsec13} and \eqref{eqsec14} in which the hydro fields are written in form of  the relations \eqref{eqsec114} and \eqref{eqsec115}. Next I use of conservation laws as in the relations \eqref{eqsec11} and \eqref{eqsec12} and try to solve them. To solve these equations, I write the sets of hydro fluctuations in the Fourier bases
     \begin{align}\label{eqsec28}
     \left(\delta T, \delta \mu, \delta u^\mu \right) \to e^{-i \omega t + i k  x} \left(\delta \tilde{T}, \delta \tilde{\mu}, \delta \tilde{u}^\mu \right).
     \end{align}
     The  sign $"\,\, \tilde \,\,\, " $ refers to the momentum-space version of fluctuations. Rotational invariance permits us to choose the momentum in an arbitrary direction. In the current paper, I select the spatial direction of momentum to be aligned in the $"x"$ direction as $k_\mu = \left(\omega, k, 0, 0\right)$. Eventually, after completing all these steps, the following  matrix valued equation is appeared
     \begin{align}\label{eqsec29}
     \mathcal{M}_{a b} \delta \phi_{b}=0, \qquad \qquad \qquad a, b = 1 , \cdots 5,
     \end{align} 
     Hydro modes are nothing but the small wave number limit of the following equation
     \begin{align}\label{eqsec210}
     	det[\mathcal{M}]=0.
     \end{align}
     Our purpose in the current section is to solve the latter equation for dense fluids. To this purpose, I have to be careful about some issues. Due to the given sets of thermo fields in the relation \eqref{eqsec27}, the scalar and vector bases in hydro fields are chosen as it follows
	 \begin{align}\label{eqsec211}
	 	s_1 &= {u^\alpha \partial_\alpha \mu \over \mu}, \qquad  s_2 = \partial \cdot u, \qquad s_3 = u^\alpha \partial_\alpha \left({T \over \mu}\right),\\\label{eqsec212}
	 	v_1^\mu &= u^\alpha \partial_\alpha u^\mu, \qquad v_2^\mu = {\Delta^{\mu \alpha} \partial_\alpha \mu \over \mu}, \qquad v_3^\mu = \Delta^{\mu \alpha} \partial_\alpha \left({T  \over \mu}\right). 
	 \end{align} 
	 I can transform these bases to the aforementioned bases in the relations \eqref{eqsec116} and \eqref{eqsec117}  by only a linear transformation. These bases are appropriate for the case of dense fluid. In order to analyze better the conditions and throughout this paper, I take the conformal symmetry to be imposed on the theory. Choosing this symmetry is not so incidental, since a dense fluid which has massless particles as its underlying theory, has conformal symmetry (if I ignore the quantum fluctuations). In practice, the conformal symmetry constrains the transport parameters in the relations  \eqref{eqsec116} and \eqref{eqsec117} and it has reduced the number of independent transport parameters \cite{Kovtun:2019hdm}. By imposing this symmetry in the four dimension of space-time, one can show that the following relations are hold between the transport parameters
	 \begin{align}\label{eqsec213}
	 	\epsilon_1= 3 \epsilon_2, \quad \epsilon_i = 3 \pi_i,\quad \nu_1 = 3 \nu_2,\quad \theta_1= \theta_2, \quad \gamma_1= \gamma_2.
	 \end{align}
	 This makes easy our job, since the number of independent transport parameter has reduced from sixteen to only nine parameters $\left(\epsilon_{1, 3}, \nu_{1, 3}, \theta_{1,3}, \gamma_{1,3}, \eta\right)$. In the current section and also in the next section, for the sake of convenience, I replace the indices "3" in parameters by index "2" and I are warned that indices $"2"$ refer to the bases $ u^\alpha \partial_\alpha \left({\mu \over T}\right)$ and $ \Delta^{\mu \alpha}\partial_\alpha \left({\mu \over T}\right)$ or $ u^\alpha \partial_\alpha \left({T \over \mu}\right)$ and $ \Delta^{\mu \alpha}\partial_\alpha \left({T \over \mu}\right)$. By using the bases shown in the relations \eqref{eqsec211} and \eqref{eqsec212}, I repeat all the aforementioned steps for deriving the hydro modes. All the linearized equations are collected together to find the matrix $\mathcal{M}_{a b}$. In what follows, I write the resultant matrix 	 
	 \begin{align}\label{eqsec214}
	 \resizebox{.9\hsize}{!}{$
	 \mathcal{M}_{a b}= 	\begin{bmatrix}
	 	-{\gamma_1 k^2 +  \omega \left(3\nu_1 \omega + i \mu_0 \chi \right)\over \mu_0} & - {\gamma_2 k^2 +\nu_2 \omega^2 \over \mu_0}& k \left(i n_0 + \left(\gamma_1 + \nu_1\right)\omega\right) & 0 &0 \\
	 	- {\theta_1 k^2 + 3 \omega \left(\epsilon_1 \omega + i n_0 \mu_0\right) \over \mu_0} & -{\theta_2 k^2 +\epsilon_2 \omega^2 \over \mu_0}&   k \left(i w_0 + \left(\epsilon_1 + \theta_1\right)\omega\right) & 0 &0 \\
	 	{k \left(i n_0 \mu_0 +\omega (\epsilon_1 + \theta_1)\right) \over \mu_0}& {k \omega \left(\epsilon_2+3\theta_2\right) \over 3 \mu_0}&- {(\epsilon_1-4\eta) k^2+ 3 \omega \left(i w_0 + \theta_1 \omega\right)\over 3} &  0 &0  \\
	 	0 & 0 &0& \eta k^2 - \omega \left(i w_0 + \theta_1 \omega\right)& 0\\
	 	0 & 0 &0&0 &\eta k^2 - \omega \left(i w_0 + \theta_1 \omega\right)
	 	\end{bmatrix}
	 	$}.
	 \end{align} 
	  In order to derive the components of this matrix, I use of the following thermo relations
	  \begin{align}\label{eqsec215}
	  {\partial p_{id} \over \partial \mu}= n_0, \qquad 	{\partial n \over \partial \mu} = \chi, \qquad  w_0 = \epsilon_0 + p_{id}, \qquad {\partial \epsilon_0 \over \partial \mu} = 	{\partial \epsilon_0 \over \partial p_{id}} {\partial p_{id} \over \partial \mu} = {n_0\over c_s^2}= 3 n_0.
	  \end{align}
	 For the matrix shown in the relation \eqref{eqsec214}, the hydro modes are derived from the solution of the following equations
	 \begin{align}\label{eqsec216}
	 	& \mbox{det} \begin{bmatrix}
	  \eta k^2 - \omega \left(i w_0 + \theta_1 \omega\right)& 0\\
	 0 &\eta k^2 - \omega \left(i w_0 + \theta_1 \omega\right)
	 	\end{bmatrix}= \left(\eta k^2 - \omega \left(i w_0 + \theta_1 \omega\right)\right)^2= 0,\\\label{eqsec217}
	 	& \mbox{det} \begin{bmatrix}
	 	-{\gamma_1 k^2 +  \omega \left(3\nu_1 \omega + i \mu_0 \chi \right)\over \mu_0} & - {\gamma_2 k^2 +\nu_2 \omega^2 \over \mu_0}& k \left(i n_0 + \left(\gamma_1 + \nu_1\right)\omega\right) \\
	 	- {\theta_1 k^2 + 3 \omega \left(\epsilon_1 \omega + i n_0 \mu_0\right) \over \mu_0} & -{\theta_2 k^2 +\epsilon_2 \omega^2 \over \mu_0}&   k \left(i w_0 + \left(\epsilon_1 + \theta_1\right)\omega\right) \\
	 	{k \left(i n_0 \mu_0 +\omega (\epsilon_1 + \theta_1)\right) \over \mu_0}& {k \omega \left(\epsilon_2+3\theta_2\right) \over 3 \mu_0}&- {(\epsilon_1-4\eta) k^2+ 3 \omega \left(i w_0 + \theta_1 \omega\right)\over 3}
	 	\end{bmatrix}=0.
	 \end{align}
	 Splitting the hydro modes into two separate equations has a physical meaning. Solutions of equation \eqref{eqsec216} are known as the shear modes, since they correspond to the perpendicular directions of fluctuations to the chosen spatial momentum. Solutions of equation \eqref{eqsec217} are known as the sound modes and they correspond to the parallel direction of fluctuations to the chosen spatial momentum . In the following subsections, I shall study these two channels, thoroughly and separately.
	 \subsection{Shear Modes} 
\begin{figure}
		\includegraphics[scale=0.8]{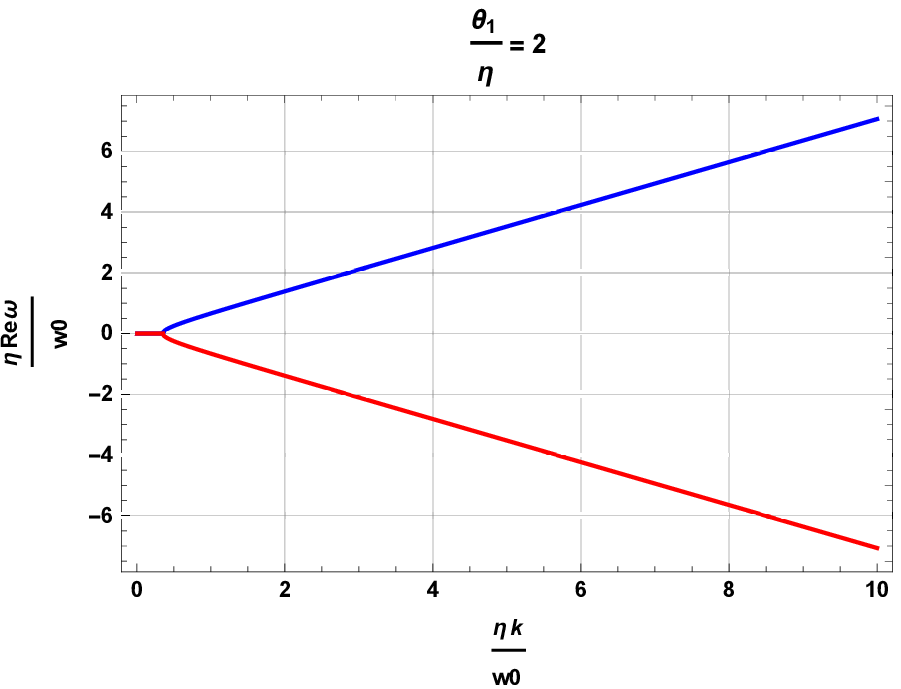}
		\hspace{0.6cm}
		\includegraphics[scale=0.8]{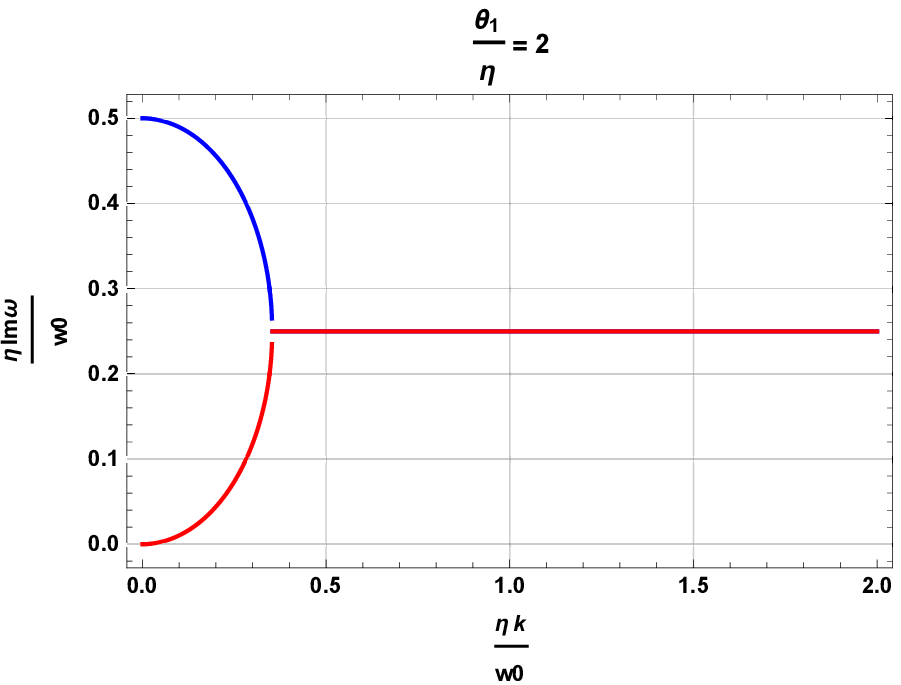}
			\caption{Real and imaginary parts of hydro modes in the shear channel for ${\theta_1 \over \eta} =2$. Left figure corresponds to the real part and right figure corresponds to the imaginary part of shear modes. Vertical and horizontal axes are relabeled so as to give dimensionless quantities. The blue and red curves represent two branches of solutions in the relation \eqref{eqsec215}.}\label{fig1}
	   \end{figure}

	 In our system, shear channels are denoted by solutions of equation \eqref{eqsec216}. The solutions are written in the following line 
	 \begin{align}\label{eqsec218}
	 	\omega = {w_0 \over 2 \theta_1} \left( i \pm \sqrt{{4 \eta \theta_1 k^2 \over w_0^2} - 1}\right).
	 \end{align}
	 Hydro modes of this channel are derived from small wavenumber limit of the latter equation
	 \begin{align}\label{eqsec219}
	 	\omega_{hydro}^{(1)} = {i w_0 \over  \theta_1} \left(1 - { \eta \theta_1 k^2 \over w_0^2}\right) + \mathcal{O}(k^4), \qquad \omega_{hydro}^{(2)} = {i \eta k^2 \over  w_0} + \mathcal{O}(k^4).
	 \end{align}
	  It is noteworthy that equation \eqref{eqsec218} transforms to the $F_{shear}$ in the paper \cite{Kovtun:2019hdm} if we set there $\textbf{v}_0=0$ and $\theta \to \theta_1$. By looking to the equation \eqref{eqsec218}, we see that there is a critical wave number
	\begin{align}\label{eqsec220}
		k_c = {w_0 \over 2 \sqrt{\eta \theta_1}},
	\end{align}
	which for $k\geq k_c$ we have two propagating modes, while for $k < k_c$ we have two nonpropagating modes \cite{Pu:2009fj}. In this channel there is one difference between our results and those of  uncharged fluid. According to the relations \eqref{eqsec216} and \eqref{eqsec218}, we have four shear modes for charged medium, while in the uncharged case we have only two shear modes. In the Fig.(\ref{fig1}) the real and imaginary parts of the hydro modes in the shear channel are sketched for ${\theta_1 \over \eta}=2$.
	Axes are relabeled so  as to give  dimensionless parameters. As we see in this figure, there is a critical wave number which before it, the real part is zero and imaginary part is nonzero. But after the $k_c$ a real part is developed for shear modes. This is the generic feature of shear modes. In the Fig.(\ref{fig1}) the critical value of momentum is ${\eta k_c \over w_0} = {1\over 2\sqrt{2}}$.\\
	Stability and causality requirements may constrain the transport parameters $\left(\eta, \theta_1\right)$. Stability requires that $Im \omega \leq 0$ and causality implies that group velocity should not exceed than 1	$\left(v_g = {\partial Re (\omega)\over \partial k} \leq 1\right)$. One might think that these two concepts are independent of each other, but I shall argue that in the relativistic theory these two issues are correlated to each other. I have observed that for our case stability gives causality.\\
	 To derive the stability constraints, I plug $\omega = i \Omega$ into the relation \eqref{eqsec216}. 	Therefore, the stability  demands that $Re \, \Omega \leq 0$. This gives rise to the following equation
	\begin{align}\label{eqsec221}
		\Omega^2 \theta_1 + \Omega w_0 + \eta k^2 =0,
	\end{align}
    which by using the Routh-Hurwitz criteria \cite{Gradshteyn:2007}, it leads to the following conditions
	\begin{align}\label{eqsec222}
		\eta \geq 0, \qquad \theta_1 \geq 0.
	\end{align}
	To derive the causality constrains, I have to look for the asymptotic limit of wave number $(k \to \infty)$ in the group velocity expression \cite{Pu:2009fj}. This statement is verified for the calculations of MIS  theory \cite{Pu:2009fj} and I can safely apply it for this general first order hydro. This is because the arguments as explained, are general and do not depend on the detail of theory. Let us  discuss briefly the main points of this argument mentioned in the paper \cite{Pu:2009fj}.  The analysis of the hydro modes in the second order dissipative hydrodynamics  went back to the original paper \cite{Hiscock:1983zz}. In this way one might guess that the MIS theory becomes always stable and causal. But this is a wrong conclusion. It was shown in the paper \cite{Denicol:2008ha} that for certain values of parameter space in the MIS formalism the theory exhibits acausal and unstable modes. Later on in the paper \cite{Pu:2009fj}, the same calculation is done for MIS formalism in an arbitrary dimension  and the authors derived the conditions which stated that MIS formalism becomes stable and causal if it respects to the following asymptotic condition\cite{Pu:2009fj}
	\begin{align}\label{eqsec223}
		\lim\limits_{k \to \infty} v_g \leq 1.
	\end{align}
	This condition states that in a general equilibrium state, the theory is stable if this asymptotic causality condition is fulfilled. The reverse is in general not true, but the stability of theory is contingent upon whether the asymptotic causality condition is satisfied. Suppose that I could write the hydro perturbations as it follows
	\begin{align}\label{eqsec224}
		\delta \mathcal{A}(x, t) = \sum_\ell \int \,\, d\omega\,  \widetilde{\delta \mathcal{A}}(\omega) e^{-i \omega t + i k_\ell(\omega) x},
	\end{align}
	 where $\delta \mathcal{A}(x, t)$ represents the hydro fields such as $\left(\delta T, \delta \mu, \delta u^\mu\right)$ and the index $"\ell"$ stands for different modes. The function $k_\ell(\omega)$ is the inverted form of dispersion relation $\omega_\ell(k)$ of the respective modes. I can set the initial conditions in such a way that $\sum_\ell \widetilde{\delta \mathcal{A}}(\omega)$ be an analytic function in the upper half section of the complex $"\omega"$ plane. Divergence of the group velocity corresponds to singularities in the complex $"\omega"$ plane. If the asymptotic causality condition \eqref{eqsec223} is respected, then the imaginary part of the dispersion relation remains always negative. Therefore, the system becomes also stable. On the other hand,  if this asymptotic condition is violated, then the singularity might appear on the upper half plane. Thus, the theory becomes unstable. To further explore the asymptotic condition \eqref{eqsec223}, I am going to calculate the equation \eqref{eqsec224}. Assume that at large $k$,  the group velocity can be written as it follows
	 \begin{align}\label{eqsec225}
	 	\lim\limits_{k \to \infty} Re \,\omega_\ell(k) = v^{as}_\ell k.
	 \end{align}
	 Thus, in this limit the exponential term in the equation \eqref{eqsec224} takes the following form
	 \begin{align}\label{eqsec226}
	 	e^{-i \omega t + i k_\ell(\omega) x} \to e^{-i {\omega \over v^{as}_\ell} \left( v^{as}_\ell t - x\right)}.
	 \end{align} 
	 For $x > v^{as}_\ell t$, I have to close the integral contour in the equation \eqref{eqsec224} in the upper half plane. If the asymptotic condition \eqref{eqsec223} is fulfilled, there are no singularities in the upper half plane and the integral gives zero. But for $x < v^{as}_\ell t$ the contour has to closed in the lower half plane and it might give the non-vanishing values if the asymptotic condition is respected. However, if I choose $v^{as}_\ell \leq 1$ (the asymptotic condition), then the position $x$ where the disturbances lie within, remains always in the light cone region. Therefore, asymptotic condition of group velocity guarantees not only the stability of theory, but also the causality of theory as a whole.\\
	 Based on this proof, to invoke the causality conditions I look for the limit $k\to \infty$ of velocity of the respective modes. For the shear channel based on the solution \eqref{eqsec218}, I get the following result for asymptotic velocity
	\begin{align}\label{eqsec227}
		\lim\limits_{k \to \infty} v_g =  \sqrt{\eta \over \theta_1}.
	\end{align}
	Thus, we conclude that causality shrinks the phase space of transport parameters to the following one
	\begin{align}\label{eqsec228}
		\eta \leq \theta_1.
	\end{align} 
	This constraint is in complete agreement with the condition (29) of the paper \cite{Kovtun:2019hdm}. In the paper \cite{Kovtun:2019hdm}, the author has obtained the latter constraint for the locally boosted fluid and it is not clear how to obtain this condition for the fluid at its  local rest frame $\textbf{v}_0=0$. The asymptotic causality condition \eqref{eqsec223} has derived the same result  without considering the velocity of frame. Also, the condition \eqref{eqsec228} can be compared with the similar condition derived for MIS fluid \cite{Pu:2009fj}. Both of these two models (MIS and general first order hydro) have shown that in a causal theory, the shear transport coefficient can not take any arbitrary value. It should take the values according to the conditions derived from equation \eqref{eqsec223}.
	\subsection{Sound Modes}
\begin{figure}
	   	\includegraphics[scale=0.8]{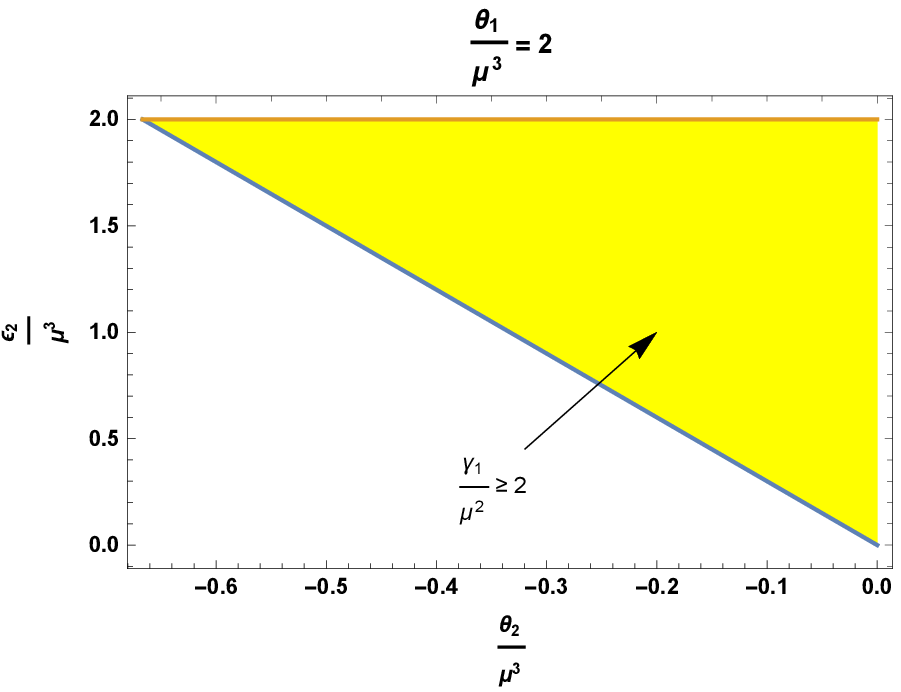}
	   	\hspace{0.6cm}
	   	\includegraphics[scale=0.8]{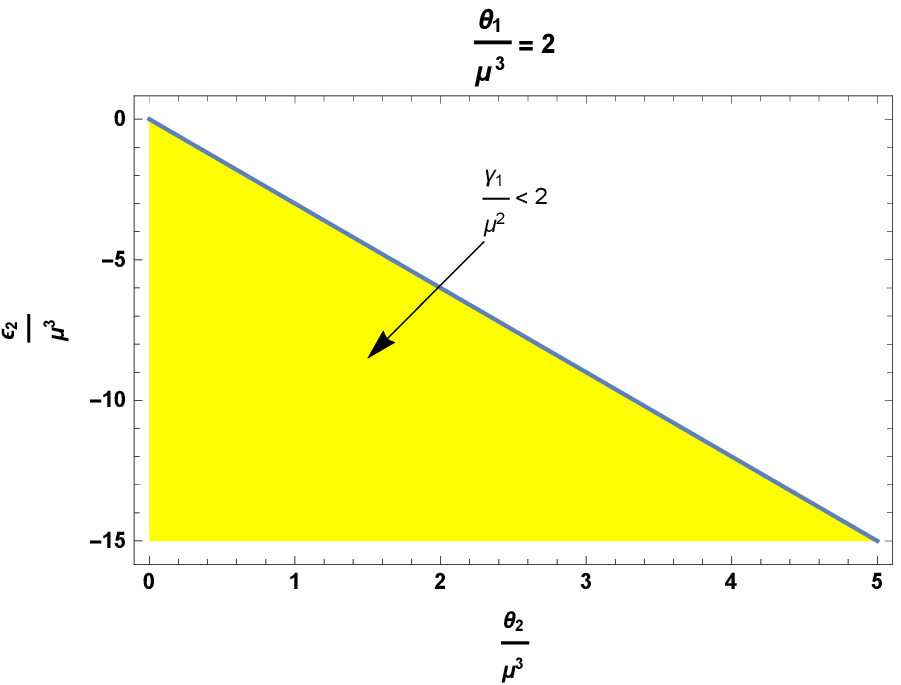}
	   		\caption{ Phase space of transports $\tilde{\epsilon}_{2}$ and $\tilde{\theta}_{2}$ for $\tilde{\theta}_1=2$. The left figure corresponds to the region bounded between $\tilde{\epsilon}_2 <2$, $\tilde{\epsilon}_2 = -3 \, \tilde{\theta}_2$ and $ \tilde{\theta}_2=0$. In this zone $\tilde{\gamma}_1\geq 2$ and $ -{2\over3}< \tilde{\theta}_2<0$. The right figure is for the case which $\tilde{\gamma}_1<2$ and $\tilde{\theta}_1 =2$. In this region $\tilde{\epsilon}_2 <0, \tilde{\theta}_2>0$ and $\tilde{\epsilon}_2 < -3 \, \tilde{\theta}_2$.}\label{fig2}
	   \end{figure}

\begin{center}
   \begin{figure}
   	\includegraphics[scale=0.8]{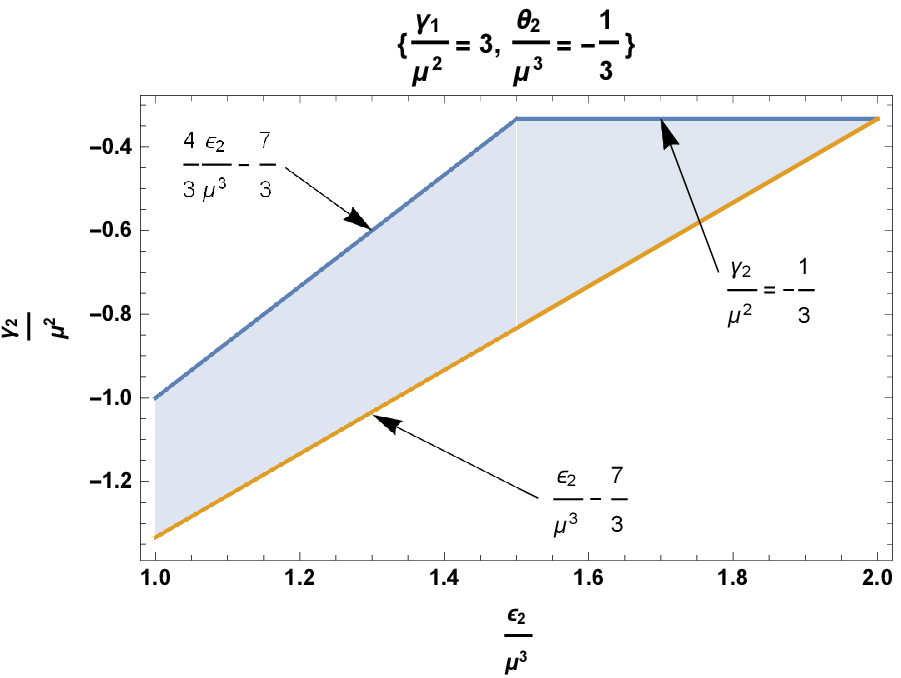}
   	\hspace{0.7cm}
   	\includegraphics[scale=0.8]{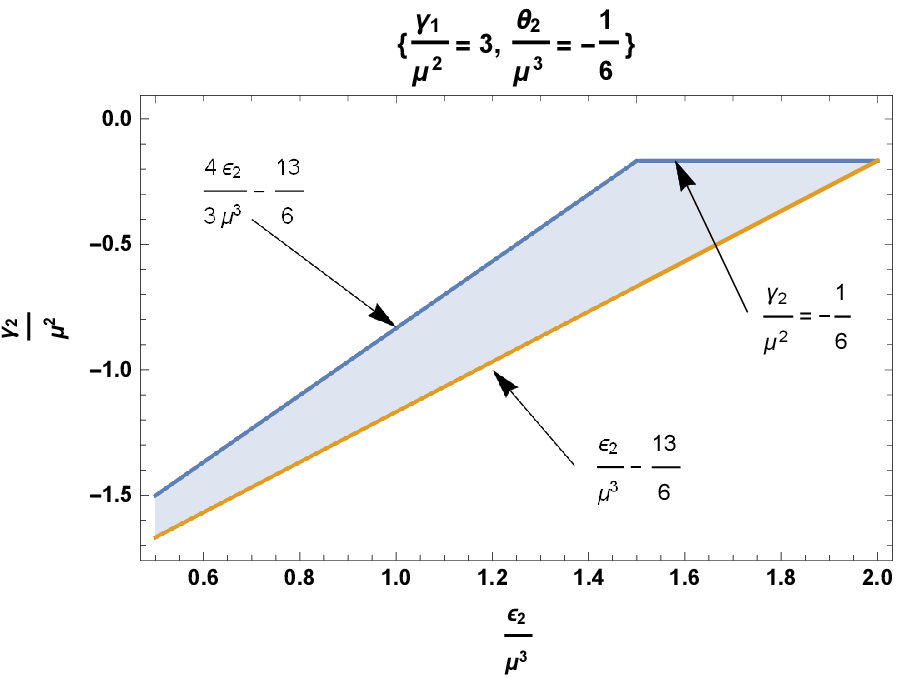}
   	\caption{Phase space of transports $\tilde{\gamma}_2$ and $\tilde{\epsilon}_2$. The left plot is for $\tilde{\gamma}_1 =3, \tilde{\theta}_2 = -{1\over 3}$ and the right plot is for $\tilde{\gamma}_1 =3, \tilde{\theta}_2 = -{1\over 6}$. In both of these plots, we have $\tilde{\gamma}_2 \leq 0$ which is compatible with the second law of thermodynamics. In each of these diagrams the functionality of boundaries are shown which are derived from Routh-Hurwitz and stability criteria. The area of left plot is 0.91 and the area of right plot is 1.04.}\label{fig3}
   \end{figure}
\end{center}

\begin{center}
    	\begin{figure}
    	\includegraphics[scale=0.85]{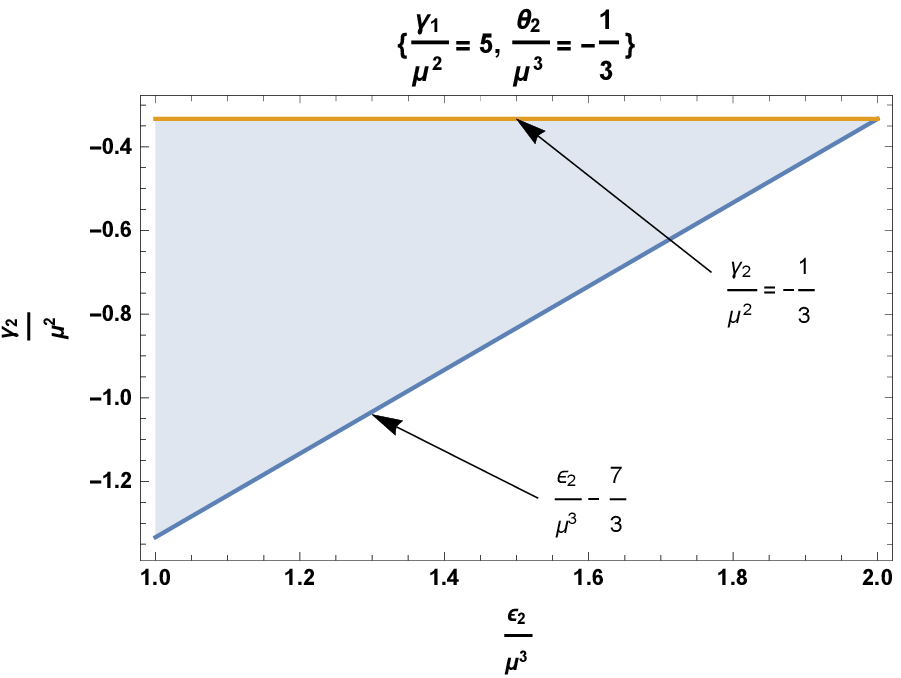}
    	\hspace{0.6cm}
    	\includegraphics[scale=0.82]{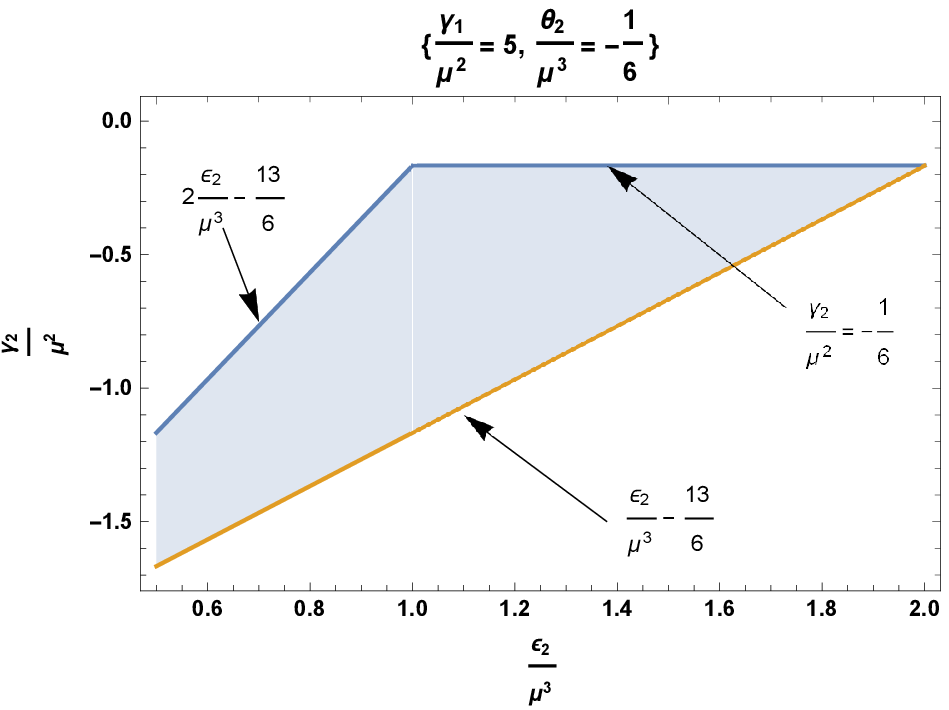}
    	\caption{Phase space of transports $\tilde{\gamma}_2$ and $\tilde{\epsilon}_2$. The left plot is for $\tilde{\gamma}_1 =5, \tilde{\theta}_2 = -{1\over 3}$ and the right plot is for $\tilde{\gamma}_1 =5, \tilde{\theta}_2 = -{1\over 6}$. In both of these plots, we have $\tilde{\gamma}_2 \leq 0$ which is compatible with the second law of thermodynamics. In each of these diagrams the functionality of boundaries are derived from Routh-Hurwitz and stability criteria. The area of left plot is 0.5 and the area of right plot is 0.875.}\label{fig4}
    	\end{figure}
    \end{center}

\begin{center}
   	\begin{figure}
   		\includegraphics[scale=0.85]{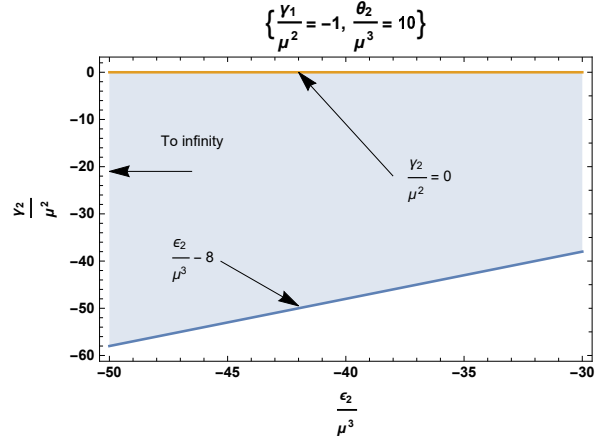}
   		\caption{Phase space of transports $\tilde{\gamma}_2$ and $\tilde{\epsilon}_2$ for $\tilde{\gamma}_1 =-1$ and $\tilde{\theta}_2 = 10$. In  this plot, we have $\tilde{\gamma}_2 \leq 0$ which is compatible with the second law of thermodynamics. The functionality of boundaries are shown which are derived from Routh-Hurwitz and stability demands.}\label{fig5}
   	\end{figure}
   \end{center}

     In our system, the sound channel is described by the solutions of equation \eqref{eqsec217}. I write the resulting equation as the following one
	\begin{align}\label{eqsec229}
		a_6 \,\omega^6 + i a_5 \,\omega^5 + a_4 \,\omega^4 + i a_3 \,\omega^3 + a_2 \,\omega^2 + i a_1 \,\omega + a_0=0.
	\end{align}
	Hydro modes correspond to the small wave number limit of the solutions of latter equation. Forms of the coefficients $(a_i,\,\, i= 1 , \ldots 6)$ in the equation \eqref{eqsec229} are given below
	 \begin{align}\label{eqsec230}
	 	a_6&\equiv 3 \theta_1 \mathcal{S}_1,\\\label{eqsec231}
	 	a_5&= \mu_0 \bigg(3  n_0 \mathcal{S}_1  - \theta_1 \mathcal{S}_2 \bigg),\\\label{eqsec232}
	 	a_4&= k^2 \bigg(3 \epsilon_1 \mathcal{S}_3 - 2 \mathcal{S}_1 \left(2\eta + \theta_1\right)\bigg) + \mu_0^2 n_0 \mathcal{S}_2,\\\label{eqsec233}
	 	a_3&= k^2 \mu_0 \left({\chi \over 3} \bigg(3 \theta_2 \epsilon_1 + \epsilon_2 \left(4 \eta + \theta_1\right)\bigg) - n_0  \bigg( \mathcal{S}_1 + \epsilon_2  \ell_1 +  3 \left( \theta_2 \nu_1 - \mathcal{S}_3 -  \epsilon_1 \ell_2 \right) +\nu_2 \left(4 \eta + \theta_1 \right)\bigg)\right),\\\label{eqsec234}
	 	a_2&={k^2 \over 3}\bigg(k^2 \bigg(\theta_1 \mathcal{S}_1 + 4 \eta \left(\gamma_1 \epsilon_2 + 3 \theta_2 \nu_1 - \theta_1 \nu_2\right) - 6 \epsilon_1 \left(\mathcal{S}_3 + 2 \eta \gamma_2\right)\bigg) - n_0 \mu_0^2 \bigg( \mathcal{S}_2 + 9 \ell_2 n_0\bigg)\bigg),\\\label{eqsec235}
	 	a_1&= {k^4 \mu_0\over 3} \left( \chi \theta_2 \left( 4 \eta - \epsilon_1 \right) + n_0  \bigg(\epsilon_2 \ell_1 - 3 \epsilon_1 \ell_2 +  3 \left(\theta_2 \nu_1 -\mathcal{S}_3 - 4 \eta \gamma_2\right) \bigg)\right),\\\label{eqsec236}
	 	a_0&= {k^4\over 3} \bigg(k^2 \left(\epsilon_1 - 4 \eta\right) \mathcal{S}_3 + 3 n_0^2 \mu_0^2 \ell_2\bigg).
	 \end{align}
	 In these relations the unknown expressions for $\mathcal{S}_{1, 2, 3}$ and $\ell_{1, 2}$ are defined as 
	 \begin{align}\label{eqsec237}
	 	& \mathcal{S}_1 \equiv \nu_2 \epsilon_1 - \nu_1 \epsilon_2, \qquad \mathcal{S}_2 \equiv \chi \epsilon_2 - 3 \nu_2 n_0, \qquad 	\mathcal{S}_3 \equiv \gamma_2 \theta_1 - \gamma_1 \theta_2,  \\\label{eqsec238}
	 		& \mathcal{\ell}_{1, 2} \equiv \gamma_{1, 2} - {n_0 \over w_0 }\theta_{1, 2} .
	 \end{align}
	For dense medium $w_0 = n_0 \mu_0$ and therefore $\ell_{1, 2} = \gamma_{1, 2} - {\theta_{1, 2}\over \mu_0}$. The $a_i$ coefficients in the relations \eqref{eqsec230} to \eqref{eqsec236} have featured some properties . The even coefficients ($a_2, a_4, a_6$) have even powers of $\mu_0$, while the odd coefficients ($a_1, a_3, a_5$) have odd powers of $\mu_0$. This is not an incidental event. It is such that the charge conjugation symmetry is satisfied. I will discuss it in few lines later. The other property is that the even coefficients $a_{2n}$ are ordered according to the even powers of momentum, while in the odd coefficients $a_{2n-1}$ there is just one term and it is proportional to the $k^{6-2n}$. The next property is that in the even coefficients the greatest powers of momentum (the order $k^6$ in $a_6$ and so on), have nothing to do with thermodynamics information and just the transport parameters appear, while the next lowest order of momentum have influenced of either the thermodynamics or transport parameters. In the odd coefficients both of  the thermodynamics values and transport parameters contribute to expressions.\\
	In the paper \cite{Kovtun:2019hdm}, the author investigated the transformation properties of transport parameters under the redefinitions of thermo fields. By looking to those transformations and extending them to  dense fluid, we are able to derive the variations of $\mathcal{S}_{1, 2, 3}$ under the fields redefinitions. The expressions $\ell_{1, 2}$ are invariant under the thermo field redefinitions. By using the equations \eqref{eqsec215}, transformation properties of transport parameters can be recast as following ones
	\begin{align}\label{eqsec239}
		&\epsilon_i \to \epsilon_i - 3 n_0 c_i, \quad \nu_i \to \nu_i - \chi c_i,\\\label{eqsec240}
		& \theta_i \to \theta_i - w_0 b_i, \quad  \gamma_i \to \gamma_i - n_0 b_i,\\\label{eqsec241}
		& \eta \to \eta.
	\end{align}
	In these relations, the ($i=1, 2$) correspond to the two sets of transport parameters. Using the latter relations will enable us to derive the transformation properties of $\mathcal{S}_{1, 2, 3}$. The final result is written as
	\begin{align}\label{eqsec242}
		&\mathcal{S}_1 \to \mathcal{S}_1 - 3 n_0 \left(\nu_1 c_2 - \nu_2 c_1\right) - \chi \left(c_1 \epsilon_2 - c_2 \epsilon_1\right),\\\label{eqsec243}
		& \mathcal{S}_2 \to \mathcal{S}_2,\\\label{eqsec244}
		& \mathcal{S}_3 \to \mathcal{S}_3 - n_0 \mu_0 \left(\ell_2 b_1 - \ell_1 b_2\right).
 	\end{align}
 Invariance of $\mathcal{S}_2$ backs to the EoS of dense fluids ($n_0= \alpha \mu_0^3, \chi = 3 \alpha \mu_0^2$) in which $\alpha$ is a positive and real number and depends on the underlying microscopic theory. In traditional view of hydrodynamics, we have fixed these free transport parameters before going to derive the hydro modes. But in the present case, I leave them free to take any values as long as the stability and causality are not violated. It is worthwhile to mention  that equations \eqref{eqsec242} to  \eqref{eqsec244} imply that $a_i$s are not invariant under the change of frame. Thus, the  stability and causality conditions derived from them, might depend on the chosen frame (special values of $b_{1, 2}$ and $c_{1, 2}$).\\
	 I back to examine the symmetries of equation \eqref{eqsec229}. One of these important  symmetries is charge conjugation symmetry which determines whether the hydrodynamic equations appear for antiparticles. If we look to the bases written in the relations \eqref{eqsec211} and \eqref{eqsec212}, we shall see that by charge conjugation transformation $(\mu_0 \to - \mu_0)$, the bases transform as the following one
	\begin{align}\label{eqsec245}
		& (s_1, \, s_2) \to (s_1, \, s_2), \quad s_3 \to - s_3,\\\label{eqsec246}
		& (v_1^\mu, \, v_2^\mu) \to (v_1^\mu, \, v_2^\mu), \quad v_3^\mu \to - v_3^\mu,\\\label{eqsec247}
		& \sigma^{\mu \nu} \to  \sigma^{\mu \nu}.
	\end{align}
	On the other hand, energy momentum tensor and charged current vector transform under the charge conjugation as it follows
	\begin{align}\label{eqsec248}
		(T^{\mu \nu}, J^\mu) \to 	(T^{\mu \nu}, -J^\mu).
	\end{align}
	Therefore, the transport parameters associated with each scalar, vector and tensor bases, change as the following one
	\begin{align}\label{eqsec249}
		& (\epsilon_1, \epsilon_2) \to (\epsilon_1, \epsilon_2), \qquad \epsilon_3 \to -\epsilon_3, \qquad \quad (\pi_1, \pi_2) \to (\pi_1, \pi_2), \quad \pi_3 \to -\pi_3, \qquad  (\nu_1, \nu_2) \to - (\nu_1, \nu_2), \quad \nu_3 \to \nu_3,\\\label{eqsec250}
		& (\theta_1, \theta_2) \to (\theta_1, \theta_2), \qquad \theta_3 \to -\theta_3, \qquad (\gamma_1, \gamma_2) \to - (\gamma_1, \gamma_2), \quad \gamma_3 \to \gamma_3,\\\label{eqsec251}
		&\eta \to \eta.
	\end{align}
     These properties give rise to the following transformations for $\mathcal{S}_{1, 2, 3}$ and $\ell_{1, 2}$ under the charge conjugation symmetry
	 \begin{align}\label{eqsec252}
	 	\left(\mathcal{S}_1, \mathcal{S}_3, \ell_2\right) \to \left(\mathcal{S}_1, \mathcal{S}_3, \ell_2\right), \qquad \left(\mathcal{S}_2, \ell_1\right) \to - \left(\mathcal{S}_2, \ell_1\right).
	 \end{align}
	 Collecting all these transformations together will result to the invariance of coefficients $\left(a_i, i =1, \cdots 6 \right)$ under the charge conjugation transformation. Therefore, the equation \eqref{eqsec229} remains invariant under the charge conjugation transformations as it is expected.\\
	 Now I gonna to derive the sound modes. Since the sound equation is a sixth order polynomial, its analytical solutions are very subtle to derive. Instead, I study them in special limits. At small wave number limit, the hydro modes can be derived as it follows
	 \begin{align}\label{eqsec253}
	 	& \omega_1 = - {i \mu_0 n_0 \over \theta_1} + \mathcal{O}(k^2),\\\label{eqsec254}
	 	& \omega_2 = {i \mu_0 \mathcal{S}_2\over 3 \mathcal{S}_1} + \mathcal{O}(k^2),\\\label{eqsec255}
	 	& \omega_{3, 4} = \pm {k \over \sqrt{3}} - {2 i k^2 \eta \left(\mathcal{S}_2 + 3 \chi \theta_2 - 9 n_0 \gamma_2\right) \over 3 n_0 \mu_0 \left(\mathcal{S}_2\ - 9 n_0 \ell_2\right)} + \mathcal{O}(k^3) = \pm {k \over \sqrt{3}} - {2 i k^2 \eta \over 3 n_0 \mu_0} + \mathcal{O}(k^3),\\\label{eqsec256}
	 	& \omega_{5, 6} = \pm k \sqrt{3 n_0 \ell_2 \over \mathcal{S}_2} + {3 i k^2 \left(9 \ell_2^2 n_0^2 a_5 + 3 n_0 \ell_2 a_3 {\mathcal{S}_2\over k^2} + a_1 {\mathcal{S}_2^2\over k^4}\right) \over 2 \mu_0^2 \mathcal{S}_2^2 n_0 \left(\mathcal{S}_2\ - 9 n_0 \ell_2\right)} + \mathcal{O}(k^3).
	 \end{align} 
	 In dense fluid the hydro modes will split into  four gapless and two gapped modes. The first two modes ($\omega_{1, 2}$) are nonpropagating modes and are about the decay of sound modes in dense fluid. They are independent of momentum (at least in the lowest order) and the transports define the new relaxation times 
	 \begin{align}\label{eqsec257}
	 \tau_1 = {\theta_1 \over \mu_0 n_0}, \qquad \tau_2 = -{3 \mathcal{S}_1 \over \mu_0 \mathcal{S}_2}.
	 \end{align}
	 Stability condition constrains the expressions in sound modes. In the $"\omega_1"$ channel,  I derive $\theta_1\geq 0$ which is nothing but the relation \eqref{eqsec222}. In the channel $\omega_2$,  I get the following result 
	 \begin{align}\label{eqsec258}
	 {\mathcal{S}_2\over \mathcal{S}_1} \leq 0
	 \end{align}
	  $\mathcal{S}_2$ is frame invariant but $\mathcal{S}_1$ is not. Therefore, changing the frame will change the condition derived from the latter equation. I can further simplify the latter constraint by using the EoS of dense fluid. This simplification leads to the
	  \begin{align}\label{eqsec259}
	  	{\epsilon_2 - \nu_2 \mu_0\over \nu_2 \epsilon_1 - \nu_1 \epsilon_2}\leq 0.
	  \end{align}
	   The channels $\omega_{3, 4}$ are the familiar sound modes which take the new modification. In the channel $\omega_{3, 4}$, stability demands that 
	  \begin{align}\label{eqsec260}
	  	{\mathcal{S}_2 + 3 \chi \theta_2 - 9 n_0 \gamma_2 \over \mathcal{S}_2\ - 9 n_0 \ell_2} \geq 0
	  \end{align}
	  By using the transformation properties shown in the relations \eqref{eqsec239} to \eqref{eqsec241} as well as the EoS of dense fluid, we could show that the latter condition is frame invariant and it reduces to the trivial condition
	  \begin{align}\label{eqsec261}
	  {\epsilon_2 - \nu_2 \mu_0 + 3 \left(\theta_2 - \gamma_2 \mu_0\right)\over \epsilon_2 - \nu_2 \mu_0 - 3 \ell_2 \mu_0} = 1\geq 0.
	  \end{align}
	  In the channel $\omega_{5, 6}$ the stability condition requires that 
	   \begin{align}\label{eqsec262}
	   	{9 \ell_2^2 n_0^2 a_5 + 3 n_0 \ell_2 a_3 {\mathcal{S}_2\over k^2} + a_1 {\mathcal{S}_2^2\over k^4} \over \mathcal{S}_2\ - 9 n_0 \ell_2} \leq 0.
	   \end{align}
	   Similar to the previous modes, I can simplify the latter result by using the EoS. The final answer is as 
	   \begin{align}\label{eqsec263}
	   	(\epsilon_2 - \nu_2 \mu_0) \mathcal{A} - 3  \ell_2 \mu_0 \mathcal{S}_1 \leq 0.
	   \end{align}
	   Expression of $\mathcal{A}$ is given in below
	   \begin{align}\label{eqsec264}
	   	& \mathcal{A} = \ell_1 \epsilon_2 + 3 \left(\theta_2 \nu_1 +\ell_2 \theta_1 - \mathcal{S}_3 - \epsilon_1 \gamma_2 \right).
	   \end{align}
	   On the other hand, in this $"\omega_{5, 6}"$ channel, the following conditions have to be satisfied 
	   \begin{align}\label{eqsec265}
	   0 \leq {3 n_0 \ell_2 \over S_2} = {\mu_0 \ell_2 \over \epsilon_2 - \nu_2 \mu_0} \leq 1.
	   \end{align}	   
	  Both of $\ell_2$ and $\mathcal{S}_2$ are frame invariant and the latter constraint is a physical constraint independent of frame redifinitions.\\
	  So far the derived conditions are from the stability requirements. I can constrain the transports from causality arguments. As argued before, the asymptotic causality criterion might give the correct result even in the boosted frame. To this purpose, I look at the large wave number limit of sound mode equation \eqref{eqsec229}. I insert $\omega \to c \, k$ in the equation \eqref{eqsec229} and pick up only the dominant terms for momentum $"k"$, because of large momentum limit. After this replacement,  the following equation is derived
	  \begin{equation}\label{eqsec266}
	  b_6 c^6 + b_4 c^4 + b_2 c^2 + b_0=0 .
	  \end{equation}
	  This equation has the following solution
	  \begin{align}\label{eqsec267}
	  	&c = \pm \sqrt{\frac{2^{4\over 3} b_4^2 -6 \times 2^{1\over 3} b_2 b_6 -2 b_4 \mathcal{B}^{1\over 3} + 2^{2\over 3} \mathcal{B}^{2\over 3}}{6 b_6 \mathcal{B}^{1\over 3}}}.
	  \end{align}
	   In the latter relation the $\mathcal{B}$ has the following definition
	   \begin{align}\label{eqsec268}
	   	&\mathcal{B} \equiv \mathcal{C} + \sqrt{\mathcal{B}^2 - 4 \left(b_4^2 -3 b_6 b_2\right)},\\\label{eqsec269}
	   	&\mathcal{C}\equiv 9 b_2 b_4 b_6 -27 b_0 b_6^2 -2 b_4^3.
	   \end{align}
	   The coefficients $\left(b_6, b_4, b_2, b_0\right)$ are leading order terms of $\left(a_6, a_4, a_2, a_0\right)$ in power of momentum $k$ which has the following form
	   \begin{align}\label{eqsec270}
	   &b_6 = a_6 = 3 \theta_1 \mathcal{S}_1,\\\label{eqsec271}
	   &b_4=  3 \epsilon_1 \mathcal{S}_3 - 2 \mathcal{S}_1 \left(2\eta + \theta_1\right),\\\label{eqsec272}
	   &b_2= {\theta_1 \mathcal{S}_1 + 4 \eta \left(\gamma_1 \epsilon_2 + 3 \theta_2 \nu_1 - \theta_1 \nu_2\right) - 6 \epsilon_1 \left(\mathcal{S}_3 + 2 \eta \gamma_2\right)\over 3} ,\\\label{eqsec273}
	   &b_0= {\mathcal{S}_3\over 3}  \left(\epsilon_1 - 4 \eta\right)  .
	   \end{align}
	   Asymptotic causality condition rules that velocity $"c"$ in the equation \eqref{eqsec267} have to be less than one. This condition together with the stability requirements shall strongly constrain the parameter space of transports. In what follows I give an example of such limitation for dense fluid.\\
	   As I have described before, the parameter space of conformal dense fluid is nine dimension. I have not any possibility to constrain all of these nine space dimension. To analyze better the phase space of transports, I choose specific values for five of them and limit the rest four according to the stability and causality requirements. These five chosen transports are
	   \begin{equation}\label{eqsec274}
	   	\tilde{\epsilon}_1 = {\epsilon_1 \over \mu_0^3} = 1, \qquad \tilde{\nu}_1 = {\nu_1 \over \mu_0^2} = 1, \qquad 
	   	\tilde{\nu}_2 = {\nu_2 \over \mu_0^2} = 2, \qquad \tilde{\eta} = {\eta \over \mu_0^3} = 1, \qquad \tilde{\theta}_1 = {\theta_1 \over \mu_0^3} = 2.
	   \end{equation} 
	   Another transports including $\left(\tilde{\epsilon}_2, \tilde{\theta}_2, \tilde{\gamma}_{1,2} = {\gamma_{1,2}\over \mu_0^2}\right)$ are limited according to the stability and causality requirements. It is worthwhile to mention here that Routh-Hurwitz  criteria will also limit the coefficients $\left(a_0, \cdots a_6\right)$ in the relations \eqref{eqsec230} to \eqref{eqsec236} as the following ones
	   \begin{align}\label{eqsec275}
	   	&\left(a_6, a_5, a_2, a_1 \right)> 0,\\\label{eqsec276}
	   	&\left(a_4, a_3, a_0 \right)< 0.
	   \end{align}
	   Therefore, I have to collect all the stability, causality and Routh-Hurwitz criteria to analyze completely the phase space of transports. The stability conditions of equations \eqref{eqsec259}, \eqref{eqsec263} and \eqref{eqsec265} as well as the Routh-Hurwitz criteria will give us the following non trivial conditions
	   \begin{align}\label{eqsec277}
	   	& \tilde{\epsilon}_2 < 2, \qquad \tilde{\gamma}_2 < \tilde{\theta}_2, \qquad \tilde{\gamma}_2 + 2 \geq \tilde{\epsilon}_2 + \tilde{\theta}_2,\\\label{eqsec278}
	   	&\mbox{If}\,\,\,\,\, \tilde{\gamma}_1 \geq 2 \to \tilde{\epsilon}_2 \geq -3 \, \tilde{\theta}_2, \quad \Rightarrow\,\,  -{2\over 3}< \tilde{\theta}_2<0,\\\label{eqsec279}
	   	&\mbox{If}\,\,\,\,\, \tilde{\gamma}_1 < 2 \to \tilde{\epsilon}_2 < -3 \, \tilde{\theta}_2, \quad \Rightarrow \,\, \tilde{\theta}_2 > 0.
	   \end{align}
	   The first line of the latter conditions is a definite condition, while the conditions in the second and third line depends on our choice. If I choose $\tilde{\gamma}_1 \geq 2$ then the space of transport $\tilde{\theta}_2$ is limited to specific values, namely between ${-2\over 3}$ and zero and if I select  $\tilde{\gamma}_1 < 2$ then the space of transport $\tilde{\theta}_2$ is all the positive real numbers. In the Fig.(\ref{fig2}) I show these limits on the phase-space of $\tilde{\epsilon}_{2}$ and $\tilde{\theta}_{2}$. These regions are solely derived from Routh-Hurwitz and stability criteria and they have nothing to do with causality. \\
   According to the arguments based on the Fig.\eqref{fig2},  I have to split the phase space of transports into two distinct regions. First the regions with $\tilde{\gamma_1} \geq 2$ and the second the regions with $\tilde{\gamma_1} <2$. I go to examine the first case, $\tilde{\gamma_1} \geq 2$.   In the Fig.\eqref{fig3} I show the valid regions of $\tilde{\gamma}_2$ and $\tilde{\epsilon}_2$ for two values $\tilde{\theta}_2 = -{1\over 3}$ and $\tilde{\theta}_2 = -{1\over 6}$ , both of them have $\tilde{\gamma_1} =3$. To derive these regions I put together all the Routh-Hurwitz, stability and causality conditions. I have observed that the conditions derived from Routh-Hurwitz and stability criteria  respect to the causality requirements. The boundary of each diagram is shown in the figure and in both of these plots $\tilde{\gamma}_2$ is negative. Form of the boundaries are derived from Routh-Hurwitz criteria as in the relations \eqref{eqsec275} and \eqref{eqsec276} and the stability conditions as those in the relations \eqref{eqsec277}. According to the regions shown in the left part of Fig.\eqref{fig2} as well as the equation \eqref{eqsec278},  I must have $1 \leq \tilde{\epsilon}_2 < 2$ for $\tilde{\theta}_2 = -{1\over 3}$ and   ${1\over 2} \leq \tilde{\epsilon}_2 < 2$ for $\tilde{\theta}_2 = -{1\over 6}$. The area of left and right plot of Fig.\eqref{fig3} are  $0.91$ and $1.04$, respectively. \vspace{-1.02cm}
    Also in the Fig.\eqref{fig4} I sketch the permissible zones of $\tilde{\gamma}_2$ and $\tilde{\epsilon}_2$ for two cases, $\tilde{\gamma}_1=5, \tilde{\theta}_2 = - {1\over 3}$ and $\tilde{\gamma}_1=5, \tilde{\theta}_2 = - {1\over 6}$. Again I have observed that Routh-Hurwitz and stability criteria respect to the causality demands.  In both of these plots $\tilde{\gamma}_2$ is negative and the boundaries are shown in each figure. The area of left and right part are 0.5 and 0.875, respectively. I have to notice that for the values $\tilde{\epsilon}_2$ and $\tilde{\theta}_2$ living on the boundaries of left part of Fig.\eqref{fig2}, there is no any acceptable region. \vspace{-1cm}
   After these two plots, I go to investigate the $\tilde{\gamma}_2 <2$. My calculations show that for $0 < \tilde{\gamma}_2 < 2$ there is no any acceptable region compatible with all requirements. For $\tilde{\gamma}_2=0$ the acceptable region is only on the line $\tilde{\gamma}_2 = {\tilde{\epsilon}_2\over 3} + \tilde{\theta}_2 -2$. For $\tilde{\gamma}_2 < 0$, an infinite acceptable region exist. In the Fig.\eqref{fig5} I show this zone for $\tilde{\gamma}_2 = -1$ and $\tilde{\theta}_2 = 10$. The boundaries are derived from Routh-Hurwitz and stability criteria. I have observed that stable regions are also in the causal region, as of the previous cases. If  the $\tilde{\theta}_2$ is increased for fixed value of $\tilde{\gamma}$, the acceptable region is between the $\tilde{\gamma}_2 = 0$ and $\tilde{\gamma}_2 = {\tilde{\epsilon}_2\over 3} + \tilde{\theta}_2 -2$. From Fig.\eqref{fig5} we could decide that the favorable region for $\tilde{\gamma}_{1,2}$ is negative values. I have mentioned here that for the values of $\tilde{\epsilon}_2$ and $\tilde{\theta}_2$ living on the boundary of right part of Fig.\eqref{fig2}, there is no any acceptable region. \vspace{-1.65cm}
   One might ask a question here and it is that do the negative values for transports seem physical or not? Our imagination (based on the second law of thermodynamics) tell us that negative values for transports are not acceptable. I know that $\eta$ and $\xi$, the shear and bulk viscosities as well as other transports are non-negative transports. Here, I discuss about theories that respect to parity and therefore I do not consider anomalous transports. we all agree about the non-negativeness of transports. Also in the first order general hydro, it has been shown that for uncharged conformal matter the on-shell solutions in the second law of thermodynamics give rise to the non-negative values for transports \cite{Kovtun:2019hdm}. However, in the charged conformal matter we shall see that negative values for transports are permissible and they do not violate the Routh-Hurwitz, stability and causality criteria. Do these negative values violate the second law of thermodynamics? Unfortunately, the answer is No!!! In the Appendix A. I show that on-shell solutions for conformal charged matter will let us to have negative values. Even for very dense medium ($\mu_0 \gg T_0$) such as in this section, the second law implies that we should have $\tilde{\gamma}_2 \leq 0$. Otherwise, the second law is violated. As we have seen form Fig.\eqref{fig5} the negative values for $\tilde{\gamma}_2$ is proved numerically.
	\section{Finite Density Medium}
	\noindent
	Finite density medium is a system which local equilibrium value of chemical potential is comparable with that of temperature $\mu_0 \sim T_0$. For this system, the hydrodynamic equations \eqref{eqsec11} and \eqref{eqsec12} can be solved consistently by setting the following combinations of thermodynamic fields and fluctuations
	\begin{align}\label{eqsec31}
	& \mbox{Thermodynamic state},\qquad T_0=T_0, \,\, \mu_0 = \mu_0, \,\, u^\mu_0= \left(1, 0,0,0\right),\\\label{eqsec32}
	& \mbox{Fluctuations},\qquad \qquad \qquad  \delta T \neq 0, \,\, \delta \mu\neq 0, \,\, \delta u^\mu= \left(0, \delta u_x, \delta u_y, \delta u_z\right).
	\end{align}
	The non vanishing amounts of $\mu_0$ and $T_0$ will permit us to choose the appropriate scalar and vector bases in agreement with our need. I let them to be as the relations \eqref{eqsec116} to \eqref{eqsec118} and derive the hydro modes by using these bases. As before, in this section I use the index number "2" instead of number "3" for transport parameters because of conformal symmetry.\\
	Our aim is to constrain these transport parameters for finite density medium. It can be done by studying the hydro modes, but by this difference that all the aforementioned conditions would depend on temperature as Ill as chemical potential. In order to derive the hydro modes, I repeat the steps given before and finally  derive the matrix $\mathcal{M}_{a b}$. Its form for finite density and temperature is written as 
	\begin{align}\label{eqsec33}
	\hspace{-0.9cm}\resizebox{1.1\hsize}{!}{$
		\mathcal{M}_{a b}= 	\begin{bmatrix}
		-{\gamma_2 k^2 +  i T_0 \beta_2 \omega + \nu_2 \omega^2 \over T_0} & - {k^2 \left(\gamma_1 T_0 - \mu_0 \gamma_2\right) + \omega \left( T_0 \left(3 \omega \nu_1 + i T_0 \beta_1\right)- \mu_0 \omega \nu_2\right)\over T_0^2}& k \left(i n_0 + \left(\gamma_1 + \nu_1\right)\omega\right) & 0 &0 \\
		- {\theta_2 k^2 + \omega \left(\epsilon_2 \omega + 3 i n_0 T_0\right) \over T_0} & - {k^2 \left(\theta_1 T_0 - \mu_0 \theta_2\right) + \omega \left( 3 i T_0 \left(w_0 - n_0 \mu_0 - i \epsilon_1 \omega\right)- \mu_0 \omega \epsilon_2\right)\over T_0^2}& k \left(i w_0 + \left(\epsilon_1 + \theta_1\right)\omega\right)& 0& 0\\
		{k \left(3 i n_0 T_0 +\omega (\epsilon_2 + 3\theta_2)\right) \over 3 T_0}& {k \left(-\mu_0 \omega \left(\epsilon_2+3\theta_2\right) + 3 T_0 \left(i w_0 - i \mu_0 n_0 + \omega \left(\epsilon_1 + \theta_1\right)\right)\right) \over 3 T_0^2}& -{1\over 3} k^2 (\epsilon_1 - 4 \eta) - \omega \left(i w_0 + \theta_1 \omega\right)& 0& 0\\
		0 & 0 &0& \eta k^2 - \omega \left(i w_0 + \theta_1 \omega\right)& 0\\
		0 & 0 &0&0 &\eta k^2 - \omega \left(i w_0 + \theta_1 \omega\right)
		\end{bmatrix}
		$}. 
	\end{align}  
	The hydro modes are nothing but the solutions of following equations
	\begin{align}\label{eqsec34}
		& \eta k^2 -\omega \left(i w_0 + \theta_1 \omega\right)=0,\\\label{eqsec35}
	&	\mbox{det} \left[\begin{array}{ccc}
			-{\gamma_2 k^2 +  i T_0 \beta_2 \omega + \nu_2 \omega^2 \over T_0} & - {k^2 \left(\gamma_1 T_0 - \mu_0 \gamma_2\right) + \omega \left( T_0 \left(3 \omega \nu_1 + i T_0 \beta_1\right)- \mu_0 \omega \nu_2\right)\over T_0^2}& k \left(i n_0 + \left(\gamma_1 + \nu_1\right)\omega\right) \\
		- {\theta_2 k^2 + \omega \left(\epsilon_2 \omega + 3 i n_0 T_0\right) \over T_0} & - {k^2 \left(\theta_1 T_0 - \mu_0 \theta_2\right) + \omega \left( 3 i T_0 \left(w_0 - n_0 \mu_0 - i \epsilon_1 \omega\right)- \mu_0 \omega \epsilon_2\right)\over T_0^2}& k \left(i w_0 + \left(\epsilon_1 + \theta_1\right)\omega\right)\\
		{k \left(3 i n_0 T_0 +\omega (\epsilon_2 + 3\theta_2)\right) \over 3 T_0}& {k \left(-\mu_0 \omega \left(\epsilon_2+3\theta_2\right) + 3 T_0 \left(i w_0 - i \mu_0 n_0 + \omega \left(\epsilon_1 + \theta_1\right)\right)\right) \over 3 T_0^2}& -{1\over 3} k^2 (\epsilon_1 - 4 \eta) - \omega \left(i w_0 + \theta_1 \omega\right)
		\end{array}
		\right]=0.
	\end{align}
	The first equation corresponds to the shear modes and the second one corresponds to the sound modes. Shear modes of this section is same as the relation \eqref{eqsec216} and all the properties, discussions and plots given before, is repeated here similarly and I do not say them again. Therefore, I analyze the sound channel.
	\subsection{Sound channel}
\begin{figure}
	\includegraphics[scale=0.85]{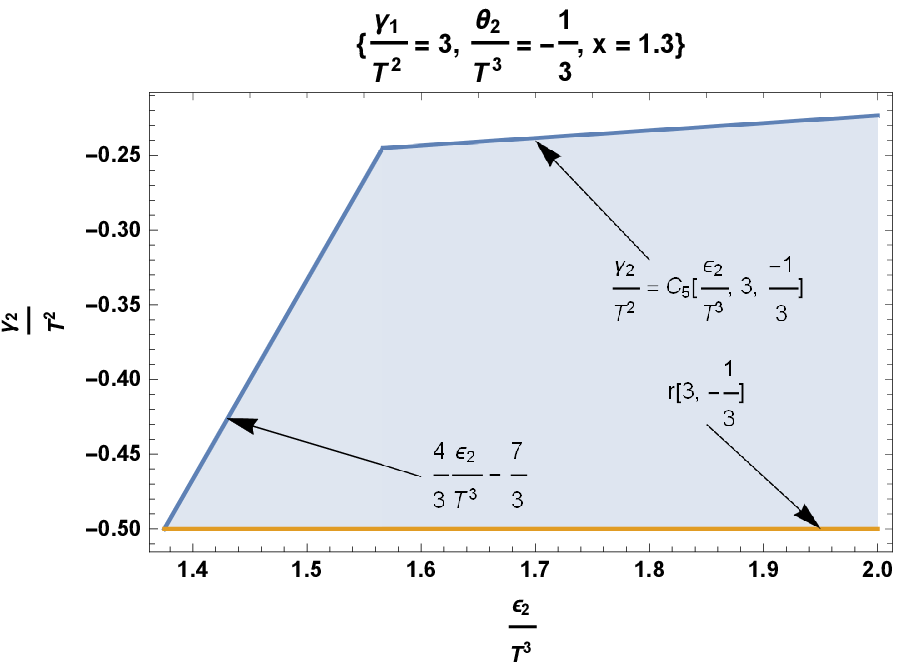}
	\hspace{0.6cm}
	\includegraphics[scale=0.85]{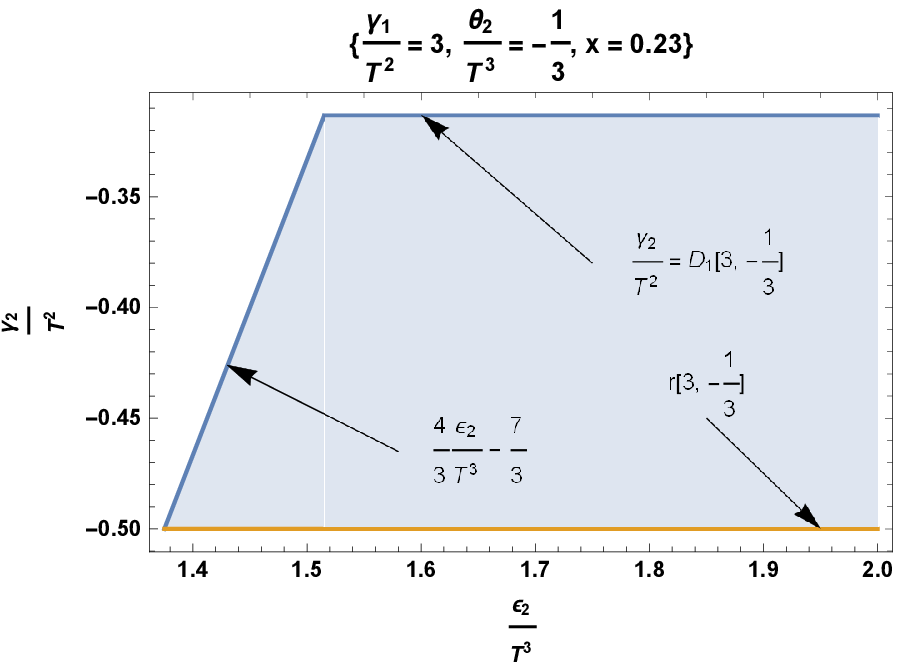}
	\hspace{0.6cm}
	\includegraphics[scale=0.85]{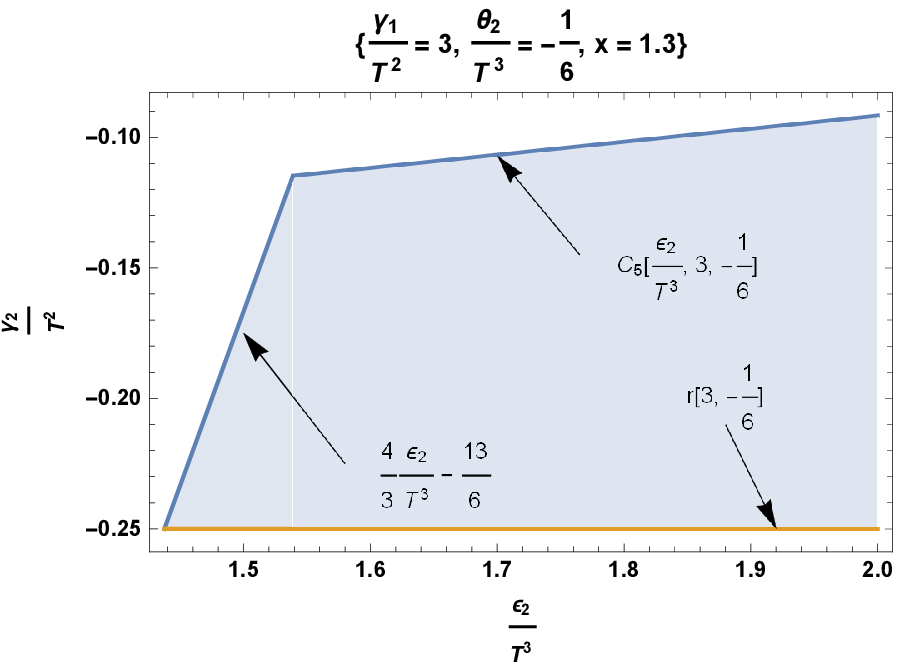}
	\hspace{0.6cm}
	\includegraphics[scale=0.85]{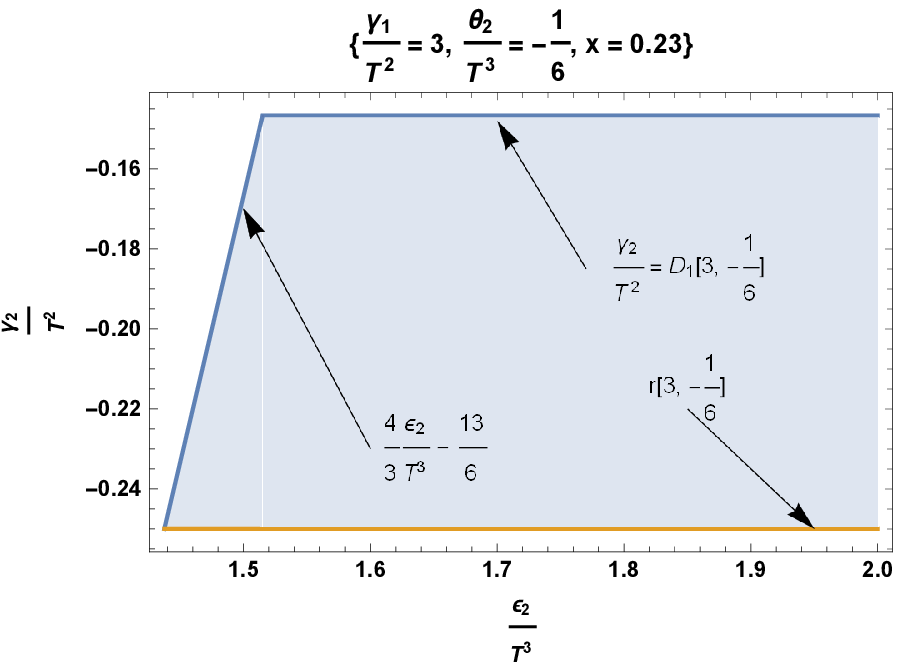}
	\caption{Acceptable zones in the case of finite density and temperature. Boundaries are given in each plot according to the stability, Routh-Hurwitz and second law constrains. Top-left figure corresponds to $\left(\tilde{\gamma}_1 = 3, \tilde{\theta}_2 = - {1\over 3}, x =1.3\right)$ and the area of this plot is 0.14. Top-right figure corresponds to $\left(\tilde{\gamma}_1 = 3, \tilde{\theta}_2 = - {1\over 3}, x =0.23\right)$ and the area of this plot is 0.10. Bottom-left figure corresponds to $\left(\tilde{\gamma}_1 = 3, \tilde{\theta}_2 = - {1\over 6}, x =1.3\right)$ and the area of this plot is 0.07. Bottom-right figure corresponds to $\left(\tilde{\gamma}_1 = 3, \tilde{\theta}_2 = - {1\over 6}, x =0.23\right)$ and the area of this plot is 0.05.}\label{fig6}
\end{figure}

\begin{figure}
	\includegraphics[scale=0.85]{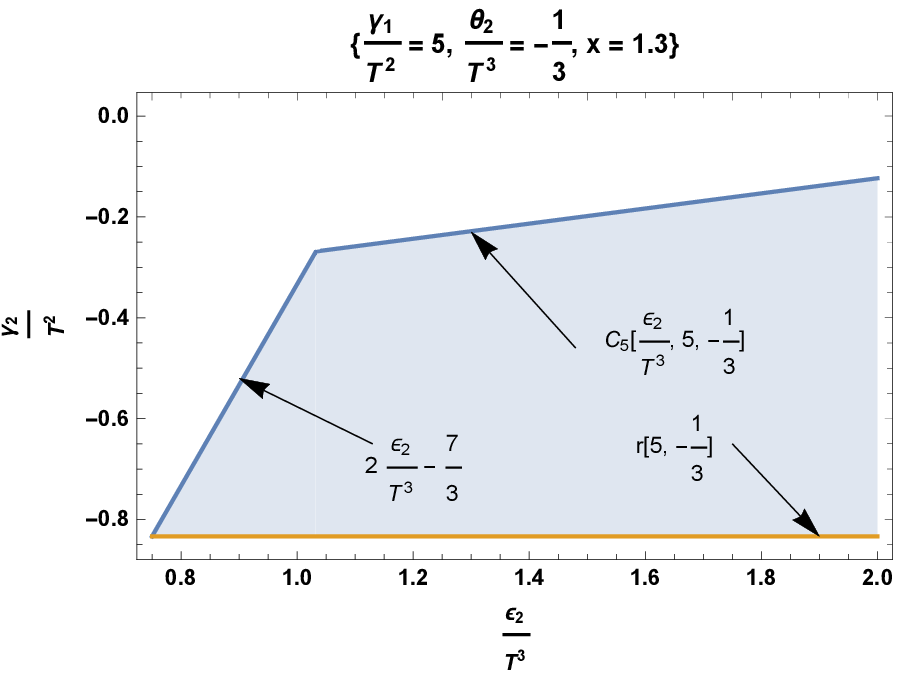}
	\hspace{0.6cm}
	\includegraphics[scale=0.85]{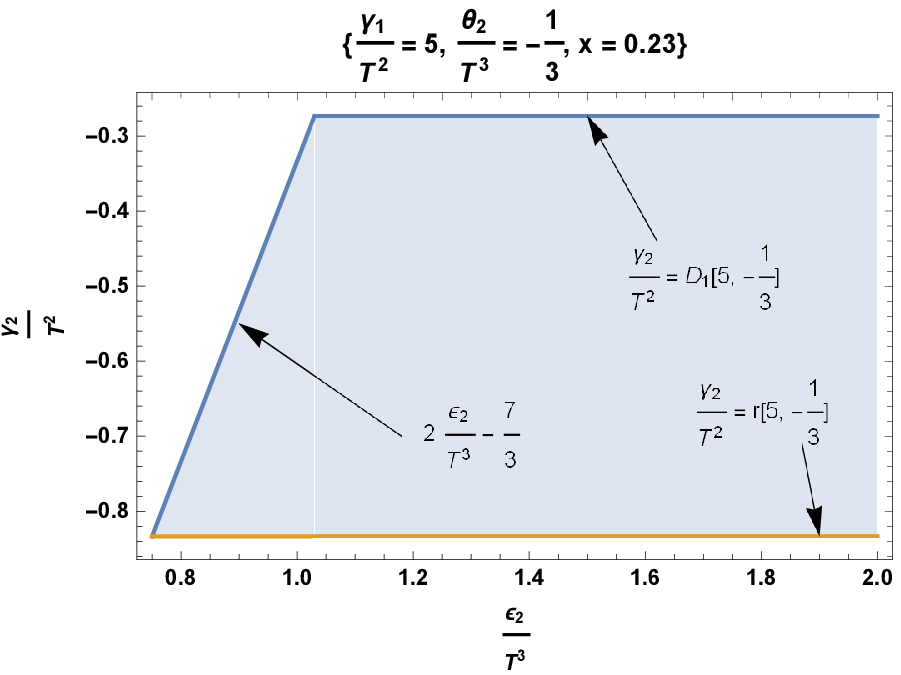}
	\hspace{0.6cm}
	\includegraphics[scale=0.85]{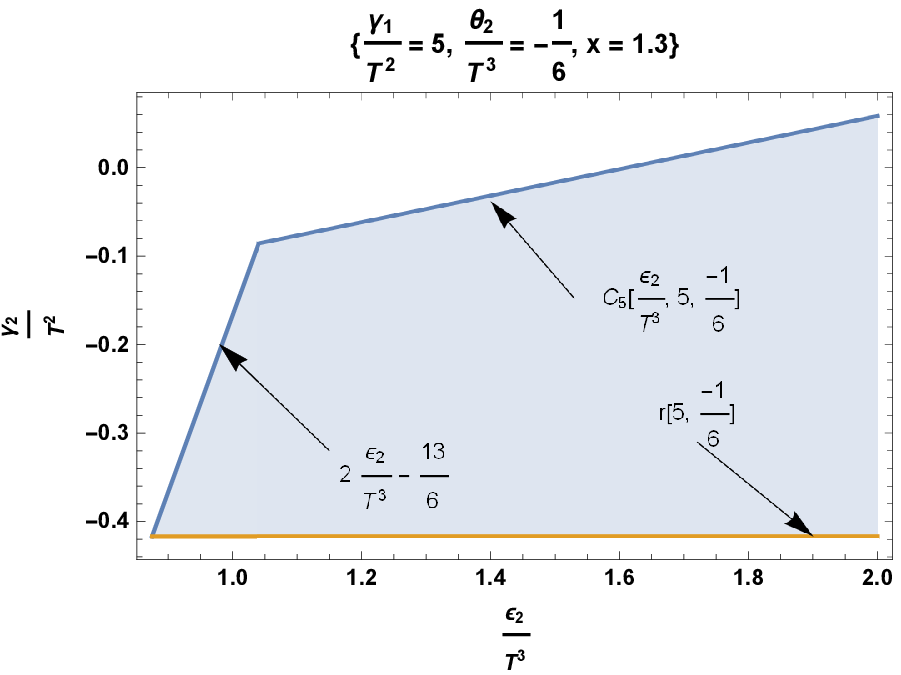}
	\hspace{0.6cm}
	\includegraphics[scale=0.85]{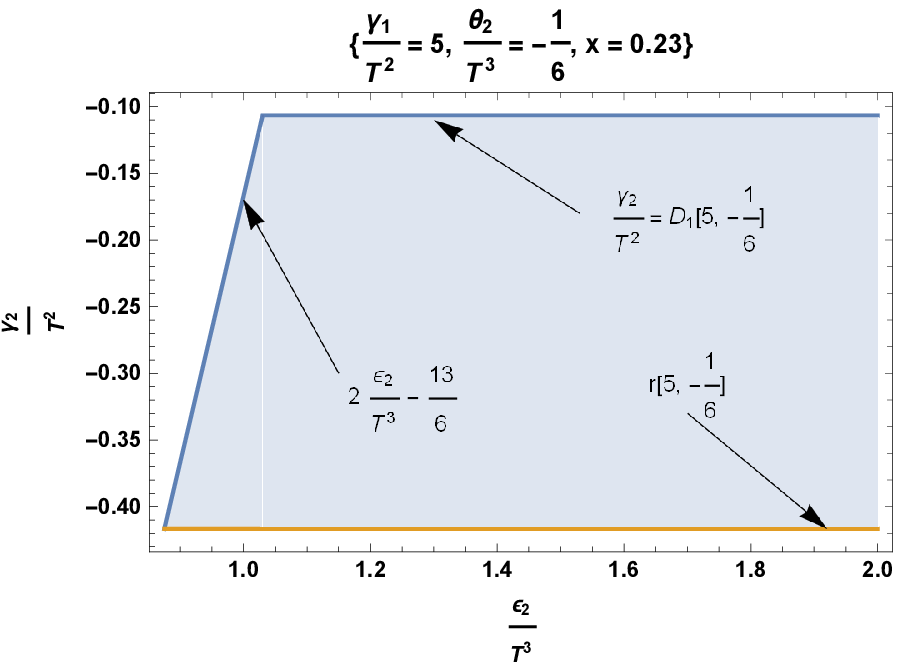}
	\caption{Acceptable zones in the case of finite density and temperature. Boundaries are given in each plot according to the stability, Routh-Hurwitz and second law constrains. Top-left figure corresponds to $\left(\tilde{\gamma}_1 = 5, \tilde{\theta}_2 = - {1\over 3}, x =1.3\right)$ and the area of this plot is 0.69. Top-right figure corresponds to $\left(\tilde{\gamma}_1 = 5, \tilde{\theta}_2 = - {1\over 3}, x =0.23\right)$ and the area of this plot is 0.61. Bottom-left figure corresponds to $\left(\tilde{\gamma}_1 = 5, \tilde{\theta}_2 = - {1\over 6}, x =1.3\right)$ and the area of this plot is 0.41. Bottom-right figure corresponds to $\left(\tilde{\gamma}_1 = 5, \tilde{\theta}_2 = - {1\over 6}, x =0.23\right)$ and the area of this plot is 0.32.}\label{fig7}
\end{figure}

\begin{figure}
	\includegraphics[scale=0.85]{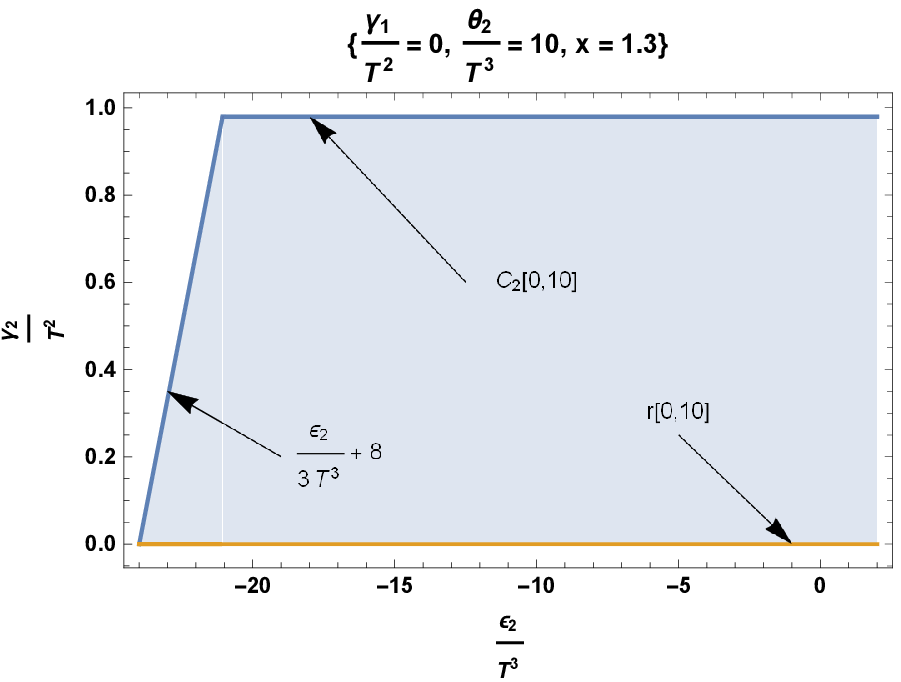}
	\hspace{0.6cm}
	\includegraphics[scale=0.85]{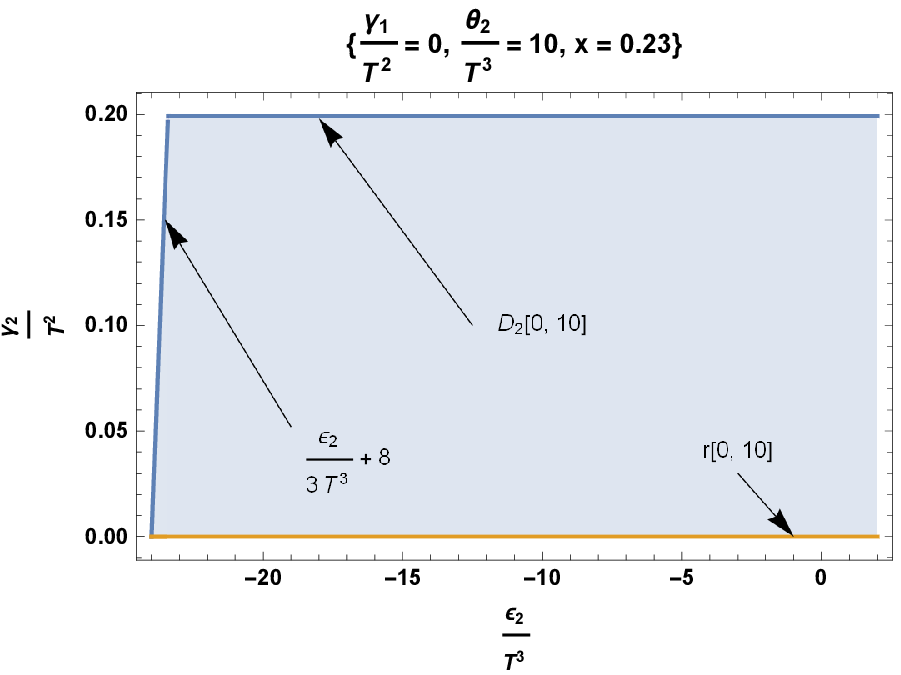}
	\caption{Acceptable zones in the case of finite density and temperature. Boundaries are given in each plot according to the stability, Routh-Hurwitz and second law constrains. The left figure corresponds to $\left(\tilde{\gamma}_1 = 0, \tilde{\theta}_2 = 10, x =1.3\right)$ and the area of this plot is 24.04. The right figure corresponds to $\left(\tilde{\gamma}_1 = 0, \tilde{\theta}_2 = 10, x =0.23\right)$ and the area of this plot is 5.11.}\label{fig8}
\end{figure}

\begin{figure}
	\includegraphics[scale=0.85]{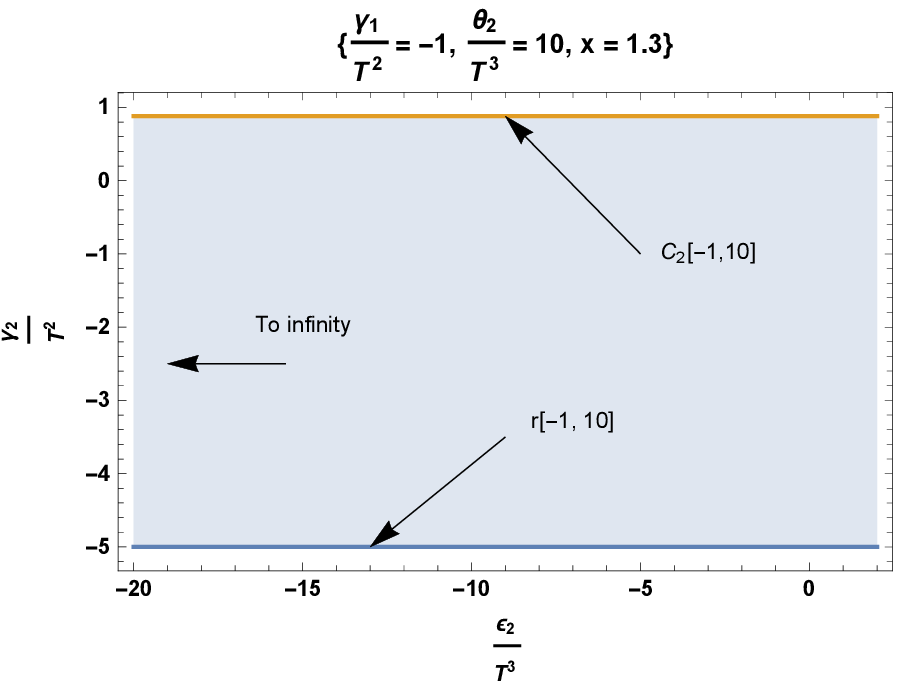}
	\hspace{0.6cm}
	\includegraphics[scale=0.85]{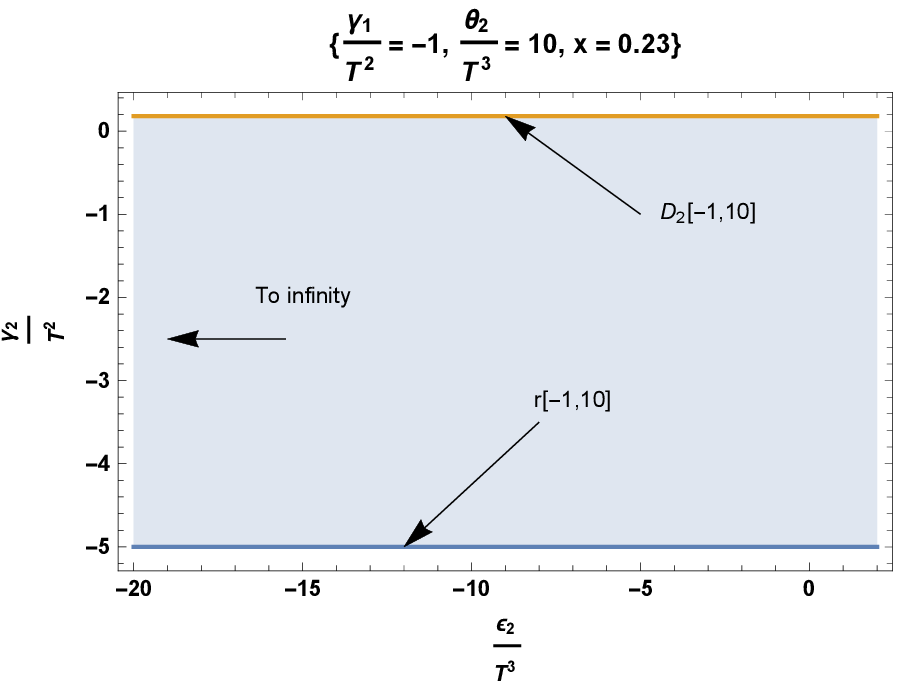}
	\caption{Acceptable zones in the case of finite density and temperature. Boundaries are given in each plot according to the stability, Routh-Hurwitz and second law constrains. The left figure corresponds to $\left(\tilde{\gamma}_1 = -1, \tilde{\theta}_2 = 10, x =1.3\right)$. The right figure corresponds to $\left(\tilde{\gamma}_1 = -1, \tilde{\theta}_2 = 10, x =0.23\right)$. For $\tilde{\gamma}_1 <0$ the accessible zone has infinite area which shows the  favorablity of this value.}\label{fig9}
\end{figure}

	\noindent
	In this channel, the equation \eqref{eqsec35} becomes
	\begin{align}\label{eqsec36}
		a_6 \,\omega^6 + i a_5 \,\omega^5 + a_4 \,\omega^4 + i a_3 \,\omega^3 + a_2 \,\omega^2 + i a_1 \,\omega + a_0=0.
	\end{align}
	The coefficients $\left(a_i, i=1, \cdots 6\right)$ for this case take the following form
	\begin{align}\label{eqsec37}
		& a_6 = 3 \theta_1 \mathcal{S}_1,\\ \label{eqsec38}
		& a_5 = 3 \bigg(w_0 \mathcal{S}_1 - \theta_1 \mathcal{S}_4\bigg),\\ \label{eqsec39}
		& a_4 = k^2 \bigg(3 \epsilon_{1} \mathcal{S}_3 - 2 \mathcal{S}_1 \left(2 \eta + \theta_{1}\right)\bigg) + 9 n_0^2 T_0 \theta_1 + 3 w_0 \mathcal{S}_4, \\ \label{eqsec310}
		& a_3 = 3 T_0 w_0 \left(3 n_0^2 - w_0 \chi \right) + k^2 \bigg(2 \left(2 \eta + \theta_{1}\right) \mathcal{S}_4 - w_0 \bigg(\gamma_1 \epsilon_{2} - 3 \gamma_{2} \epsilon_{1} + 3 \theta_{2} \nu_1 -\nu_{2} \theta_{1} + \mathcal{S}_1 - 3 \mathcal{S}_3\bigg)\bigg),\\\label{eqec311}
		& a_2 = - k^2 \bigg( w_0 \bigg(\mathcal{S}_4 + 3 w_0 \ell_2 -T_0 \left(3 n_0 \gamma_{1} + \chi \left(4 \eta + \theta_1\right)\right)\bigg) + 6 n_0^2 T_0 \left(2 \eta +\theta_{1}\right)\bigg)\nonumber\\
		& \hspace{0.65cm}+ {k^4 \over 3} \bigg(\theta_{1} \mathcal{S}_1 - 6 \epsilon_{1} \mathcal{S}_3 + 4 \eta \left(\gamma_{1} \epsilon_{2} - 3 \gamma_{2} \epsilon_{1} + 3 \theta_{2} \nu_1 - \theta_{1} \nu_{2}\right)\bigg),\\\label{eqsec312}
		& a_1 =  {k^4 \over 3} \bigg(w_0 \left(\epsilon_{2} \ell_1 - 3 \epsilon_{1} \ell_2 - 3 \left(\mathcal{S}_3 + 4\eta \gamma_{2} -\theta_{2} \nu_1 \right)  \right) -3 n_0 T_0 \left(\theta_{1} \nu_1 -4\eta \gamma_{2}\right) + \left(\epsilon_{1} - 4\eta \right) \left(\chi \theta_{1} T_0 -3 n_0 \theta_{2}\right)\bigg) \nonumber\\
		&\hspace{0.45cm} -  k^2 T_0 w_0 \left(3 n_0^2 - w_0 \chi \right),\\\label{eqsec313}
		& a_0 =  {k^4 \over 3} \bigg(k^2 \left(\epsilon_{1} - 4 \eta\right)\mathcal{S}_3 + 3 w_0^2 \ell_2 - 3 n_0 T_0 w_0 \ell_1\bigg).
	\end{align} 
	Definitions of $\mathcal{S}_{1,3}$ and $\ell_{1, 2}$ are same as the relations \eqref{eqsec237} and \eqref{eqsec238}. The $\mathcal{S}_4$ has the following definition
	\begin{align}\label{eqsec314}
		\mathcal{S}_4 \equiv \epsilon_{2} n_0 - w_0 \nu_2 + T_{0} \left(3 n_0 \nu_1 -\chi \epsilon_{1}\right).
	\end{align}
	In order to reach to the expressions for $a_i$ coefficients I use of the following equation between number density susceptibilities for conformal matter
	\begin{align}\label{eqsec315}
		 T_0  \left({\partial n_0 \over \partial T_0 }\right)_{\mu_{{0}}} + \mu_0 \left({\partial n_0 \over \partial \mu_0 }\right)_{T_{{0}}} = 3 n_0.
	\end{align}
	The $a_i$ coefficients in the relations \eqref{eqsec37} to \eqref{eqsec313} have the similar features as of the relations \eqref{eqsec230} to \eqref{eqsec236} which are described below the relation \eqref{eqsec238}. Thus, I do not repeat them again. So, I go straightly to the sound hydro modes 
	\begin{align}\label{eqsec316}
	    & \omega_{1} = - {i w_0 \over \theta_{1}} + \mathcal{O}(k^2), \\ \label{eqsec317}
	    & \omega_{2, 3} = {i \over 2 \mathcal{S}_1} \bigg(\mathcal{S}_4 \pm \sqrt{\mathcal{S}_4^2 + 4 T_0 \mathcal{S}_1 \left(3 n_0^2 - w_0 \chi \right)}\bigg)  + \mathcal{O}(k^2), \\ \label{eqsec318}
	    & \omega_{4} = {i k^2 \left(n_0 T_0 \ell_1 - w_0 \ell_2\right)\over T_0 \left(3 n_0^2 - w_0 \chi \right)} + \mathcal{O}(k^3) ,\\\label{eqsec319}
		& \omega_{5, 6} = \pm {k \over \sqrt{3}} - {2 i k^2 \eta \over 3 w_0}.		
	\end{align}
	Unlike the previous section, in the current section the sound mode possess three gapless and three gapped modes. The channels $\omega_{1, 2, 3}$ are about the decay of sound modes in this case and the corresponding relaxation times can be written as 
	\begin{align}\label{eqsec320}
		\tau_1 \equiv {\theta_{1} \over w_0}, \qquad \tau_{2, 3} \equiv {- 2 \mathcal{S}_1 \over \mathcal{S}_4 \pm \sqrt{\mathcal{S}_4^2 + 4 T_0 \mathcal{S}_1 \left(3 n_0^2 - w_0 \chi \right)}}.
	\end{align} 
	To constrain the transports we have four ways; i) Routh-Hurwitz criteria, ii) stability, iii) causality and iv) second law of thermodynamics. I have to tune the transports in such a way that all these four requirements are satisfied simultaneously. The Routh-Hurwitz criteria can be imposed independently. The stability demands that 
	\begin{align}\label{eqsec321}
		& {n_0 T_0 \ell_1 - w_0 \ell_2\over 3 n_0^2 - w_0 \chi} \leq 0,\\\label{eqsec322}
		& {\mathcal{S}_4 \pm \sqrt{\mathcal{S}_4^2 + 4 T_0 \mathcal{S}_1 \left(3 n_0^2 - w_0 \chi \right)} \over \mathcal{S}_1} \leq 0, 
	\end{align}
	as well as the relation \eqref{eqsec222}. The causality requirements have to be imposed according to the relation \eqref{eqsec223}. Since the high momentum terms of even $a_i$s in the expressions \eqref{eqsec37} to \eqref{eqsec313} are like as the even $a_i$s in the relations \eqref{eqsec230} to \eqref{eqsec236}, the asymptotic velocity given in the relation \eqref{eqsec267} can be applied similarly in this case. The only important thing in this section is that how to apply the second law requests for constraining the transports. This is not so hard problem, since the relation \eqref{eqapp20} is our guide. \\
	To be some concrete and to our analysis program, I take the EoS of weakly interacting QGP with $N_c$ gluons and $N_f$ fermions
	\begin{align}\label{eqsec323}
		\epsilon_0 = 2 (N_c^2 - 1) {\pi^2 T^4 \over 30} + 2 N_f \left({7 \pi^2 T^4 \over 120} + {\mu^2 T^2 \over 4} + {\mu^4 \over 8 \pi^2}\right).
	\end{align}
	The factor $2$ accounts for the spin DoF. In what follows, I take $N_c = N_f = 3$ throughout the analysis. Hereafter, I split our analysis to two cases: 1) $x= {\mu_0 \over T_0} = 1.30$ and 2) $x = {\mu_0 \over T_0} = 0.23$. Choice of these values are such that the variable ${4 \epsilon_0 \over 3 n_0 T_0}$ becomes $50$ and $10$, respectively. From Routh-Hurwitz and stability requirements we have two conditions independent of $x$
	\begin{align}\label{eqsec324}
		\tilde{\gamma}_{2} > r[\tilde{\gamma}_{1},  \tilde{\theta}_{2}] = {\tilde{\gamma}_{1} \tilde{\theta}_{2} \over 2}, \qquad 2 \tilde{\epsilon}_{2} \tilde{\gamma}_{1} + 3 \left(2 + \tilde{\gamma}_{1}\right) \tilde{\theta}_{2} \geq 6 + \tilde{\epsilon}_{2} + 12 \tilde{\gamma}_{2}.
	\end{align}
	Unlike the previous section, in this section I scale the transports in terms of $T_0$
	\begin{align}\label{eqsec325}
		\tilde{\epsilon}_{1,2} = {\epsilon_{1, 2}\over T^3},\qquad \tilde{\theta}_{1,2} = {\theta_{1, 2}\over T^3}, \qquad \tilde{\eta} = {\eta\over T^3}, \qquad \tilde{\gamma}_{1,2} = {\gamma_{1, 2}\over T^2}, \qquad 
		\tilde{\nu}_{1,2} = {\nu_{1, 2}\over T^2}
	\end{align} 
	Also RouthHurwitz, stability and second law give another constraints which depend on $x$. For two above mentioned values of $x$, namely the $x= 1.3$ and $x=0.23$ I have listed them in Table\eqref{table1}.
	\begin{table}
		\begin{center}	
			\hspace{-1cm}\resizebox{1.05\hsize}{!}{$\begin{tabular}{|c|c|}
				\hline
				$x = 1.3$ & $x = 0.23$\\
				\hline
				$\tilde{\gamma}_2 \leq C_1 [\tilde{\gamma}_{1},  \tilde{\theta}_{2}] = \tilde{\theta}_2 + 0.1 \tilde{\gamma}_1 -0.2$ & $\tilde{\gamma}_2 \leq D_1 [\tilde{\gamma}_{1},  \tilde{\theta}_{2}] = \tilde{\theta}_2 + 0.02 \tilde{\gamma}_1 -0.04$\\
				\hline
				$\tilde{\gamma}_2 \leq C_2 [\tilde{\gamma}_{1},  \tilde{\theta}_{2}] = 0.1 \left( \tilde{\gamma}_1 + \tilde{\theta}_2\right) -0.02$ & $\tilde{\gamma}_2 \leq D_2 [\tilde{\gamma}_{1},  \tilde{\theta}_{2}] = 0.02 \left( \tilde{\gamma}_1 + \tilde{\theta}_2\right) -0.0008$\\
				\hline
				$\tilde{\gamma}_2 \leq C_3 [\tilde{\epsilon}_2, \tilde{\gamma}_{1},  \tilde{\theta}_{2}] = 1.37 + \tilde{\epsilon}_2 \left(0.11 \tilde{\gamma}_1 - 0.2\right) + {\tilde{\theta}_2 \over 3} \left(1 + \tilde{\gamma}_1\right)$& $\tilde{\gamma}_2 \leq D_3 [\tilde{\epsilon}_2, \tilde{\gamma}_{1},  \tilde{\theta}_{2}] = 1.57 + \tilde{\epsilon}_2 \left(0.11 \tilde{\gamma}_1 - 0.12\right) + {\tilde{\theta}_2 \over 3} \left(1 + \tilde{\gamma}_1\right)$\\
				\hline
				$\tilde{\gamma}_2 \leq C_4 [\tilde{\epsilon}_2, \tilde{\gamma}_{1},  \tilde{\theta}_{2}] = 0.71 - 0.03 \tilde{\epsilon}_2 + 0.1 \tilde{\gamma}_1 + \tilde{\theta}_2$ & $\tilde{\gamma}_2 \leq D_4 [\tilde{\epsilon}_2, \tilde{\gamma}_{1},  \tilde{\theta}_{2}] = 0.84 - 0.006 \tilde{\epsilon}_2 + 0.2 \tilde{\gamma}_1 + \tilde{\theta}_2$ \\
				\hline
				$\tilde{\gamma}_2 \leq C_5 [\tilde{\epsilon}_2, \tilde{\gamma}_{1},  \tilde{\theta}_{2}] = -0.06 + \tilde{\epsilon}_2 \left(0.05 \tilde{\gamma}_1 - 0.1\right) + \tilde{\theta}_2 \left(0.15 \tilde{\gamma}_1 + 0.34\right)$ &$\tilde{\gamma}_2 \leq D_5 [\tilde{\epsilon}_2, \tilde{\gamma}_{1},  \tilde{\theta}_{2}] = -0.03 + \tilde{\epsilon}_2 \left(0.048 \tilde{\gamma}_1 - 0.096\right) + \tilde{\theta}_2 \left(0.14 \tilde{\gamma}_1 + 0.29\right)$ \\
				\hline
			\end{tabular}
			$}
		\caption{Lists of constrains for two values of $x = {\mu_0 \over T_0}$, including the $x=1.3$ in the left and $x=0.23$ in the right column, stemming from the Routh-Hurwitz, stability and second law criteria.}\label{table1}
		\end{center}
	\end{table}
In order to compare better the analysis with the similar case in the dense fluid, I choose the same values for $\left(\tilde{\epsilon}_{1,2}, \tilde{\nu}_1, \tilde{\theta}_1, \tilde{\eta}\right)$ as the ones denoted in the relation \eqref{eqsec274}. Also the chosen values for $\tilde{\gamma}_1$ and $\tilde{\theta}_2$ to draw the figures is same as before. I mean that first I split the plots into two distinct branches: first the branch with $\tilde{\gamma}_1 \geq 2$ and second the branch with $\tilde{\gamma}_1 <2$. For the first branch I take four plots with $\left(\tilde{\gamma}_1 = 3, \tilde{\theta}_2 = - {1\over 3}\right)$,  $\left(\tilde{\gamma}_1 = 3, \tilde{\theta}_2 = - {1\over 6}\right)$, $\left(\tilde{\gamma}_1 = 5, \tilde{\theta}_2 = - {1\over 3}\right)$ and  $\left(\tilde{\gamma}_1 = 5, \tilde{\theta}_2 = - {1\over 6}\right)$ for each of $x= 1.3$ and $x = 0.23$. After that I discuss about the existence of solutions for $0 < \tilde{\gamma}_1 < 2$. Eventually, I illustrate the figures for $\tilde{\gamma}_1 <0$ for each of the $x$ values. The figures show the physical $\tilde{\gamma}_2$ and $\tilde{\epsilon}_2$ zones which are compatible with all the Routh-Hurwitz, stability, causality and second law requirements. This is our strategy to know better the accessible zones of transports.\\
In the Fig.\eqref{fig6} the acceptable zones for $x=1.3$ and $x=0.23$ with the sets $\left(\tilde{\gamma}_1 = 3, \tilde{\theta}_2 = - {1\over 3}\right)$,  $\left(\tilde{\gamma}_1 = 3, \tilde{\theta}_2 = - {1\over 6}\right)$ is shown. The boundaries of each plot is indicated which are labeled by the corresponding conditions given in the Table. \eqref{table1} and the relation \eqref{eqsec325}. The areas of the plots from top-left to bottom-right are $\left(0.14, 0.10, 0.07, 0.05\right)$, horizontally. I have also seen that the values of transports inside these zones do not contradict with the causality condition, the relation \eqref{eqsec223} in which the asymptotic velocity is given by the relation \eqref{eqsec267}. Compared to the similar plots in the Fig. \eqref{fig3}, I have seen that finite ratio of $x$ for these values of $\tilde{\gamma}_1$ and $\tilde{\theta}_2$ have decreased the accessible zone. It seems that if $x \to 0$ the acceptable zone shrinks even more and more and thus for these values of $\tilde{\gamma}_1$ and $\tilde{\theta}_2$ the high density medium is much more favorable.\\
\noindent
In the Fig.\eqref{fig7} the same plots are shown for  for $x=1.3$ and $x=0.23$ with  $\left(\tilde{\gamma}_1 = 5, \tilde{\theta}_2 = - {1\over 3}\right)$,  $\left(\tilde{\gamma}_1 = 5, \tilde{\theta}_2 = - {1\over 6}\right)$. The boundaries of each plot is labeled by the corresponding conditions given in the Table.\eqref{table1} and the relation \eqref{eqsec325}. The areas of the plots from top-left to bottom-right are $\left(0.69, 0.61, 0.41, 0.32\right)$, horizontally. These physical zones have satisfied the asymptotic causality condition, given in the relation \eqref{eqsec223}. Compared to the similar plots in the Fig.\eqref{fig4}, I have seen that finite ratio of $x$ for  $\left(\tilde{\gamma}_1 = 5, \tilde{\theta}_2 = - {1\over 3}\right)$ have increased the accessible zone, while for $\left(\tilde{\gamma}_1 = 5, \tilde{\theta}_2 = - {1\over 6}\right)$ have decreased it. Again  for the values $\tilde{\epsilon}_2$ and $\tilde{\theta}_2$ living on the boundaries of left part of Fig.\eqref{fig2}, there is no any acceptable region.\\
For $0 < \tilde{\gamma}_1 < 2$ there is no any accessible zone. For $\tilde{\gamma}_1 = 0$ until a critical value of $\tilde{\theta}_2$ we have no any physical zone, but after it a large area appears. In the Fig.\eqref{fig8} I show this area for $\left(\tilde{\gamma}_1 =0, \tilde{\theta}_2 =10\right)$ for each of the $x$ values. Compared to dense medium, I have seen that the physical spaces for transport becomes larger which shows that for $\tilde{\gamma}_1 = 0$ the finite density medium is more favorable than dense medium. To remind, I have to say that in dense medium for $\tilde{\gamma}_1 = 0$, the valid space is on the line $\tilde{\gamma}_2 = {\tilde{\epsilon}_2 \over 3} + \tilde{\theta}_2 -2$. \\ 
If $\tilde{\gamma}_1 < 0$ again until a critical value of $\tilde{\theta}_2$ we have no any valid zone compatible with all conditions, but after it an infinite physical space emerges. In the Fig.\eqref{fig9} I indicate this zone for $\left(\tilde{\gamma}_1 = -1, \tilde{\theta}_2 =10\right)$ for each of the $x$ values. Unlike the all previous cases, for systems with $\tilde{\gamma}_1 < 0$ in the finite ratio of $x$ and after the critical $\tilde{\theta}_2$, the physical zone is infinite. Similar case for dense medium is shown in the Fig.\eqref{fig5} which also has infinite area. It seems such that the space $\tilde{\gamma}_2 < 0 $ is most favorable region for finite density medium.
	\section{Conclusion}
	Looking to the stability and causality problems in dissipative hydrodynamics is one of the long standing challenges in this field. Stability means that hydro fluctuations never grow up as the time runs and causality refers to the fact that the velocity of  fluctuations never exceed than speed of light. Historically, this problem is remedied by adding a phenomenological equation to the known hydro equation and treating the dissipative tensors as new DoF. This strategy is good but it suffers from lacking the fundamental bases and has no any physical background to support it. The newly developed GF  notion has resolved the stability and causality problems without introducing  artificial terms. This idea has benefited of the frame concept in dissipative hydrodynamic and does not fix it before studying the stability and causality conditions.  Indeed, the physical hydrodynamic frame for relativistic systems in view of the GF is a frame in which respects to the all physical and high energetic conditions.\\
    This work tries to give the physical and acceptable region  of transports for dense medium. The selected framework in this paper is to work with conformal charged matter in order to analyze better the conditions. I have seen that the charge conjugation symmetry implies RH equations for particles and antiparticles are the same. The main achievement of this paper is that  for conformal charged matter the second law of thermodynamics has not ruled out the existence of negative transports. Importance of this result is that so far we all agree on the non-negativeness of transports because of second law's rule $\left(\partial_{\mu} S^\mu \geq 0\right)$ and appearance of negative transports seems to be a taboo. I illustrate the existence of negative transports theoretically and numerically. The sign of scalar transports for conformal matter in the GF framework are not limited by any constraints, but there is a condition among the vector transports, i.e. the relation \eqref{eqapp20} which only tells about the sign of transports combination not the individual transports. However, I infer that in extreme limits which $x \to 0$ or $x \to \infty$, the second law of thermodynamic has ruled definitely that $\tilde{\gamma}_2 \leq 0$. By fixing some transports, the good regions for other transports including the $\tilde{\gamma}_2$ and $\tilde{\epsilon}_2$ are derived. In the case of finite $T_0$ and $\mu_0$ medium, this work is done for two ratios of $\left(x= 1.3, 0.23\right)$ and the conditions for each of these $x$ is derived and tabulated. From the areas of plot we could judge about the favorability of each case and compare it  with other ones. I have seen that the most favorable regions for transports $\tilde{\gamma}_{1, 2}$ are negative values. I have also observed that the regions which are derived from Routh-Hurwitz, stability and second law constraints, have respected to the asymptotic causality condition expressed in the relation \eqref{eqsec223}.\\
	The GF framework is in its infant age and deserves a lot attention. In following this paper we could extend it in some lines. First we have to know about the physical meaning of negative transports. I mean that we have to explore the implications of this negative transports in other applications of hydrodynamic calculations and see whether or how a pathological behavior emerges or not.  Second we can construct a microscopic relation for these transports, i.e. from the Green-Kubo formalism or other relations and to observe the fingerprints of this negativeness on the microscopic field theory. Other important question is that to construct a holographic picture for the GF hydrodynamic framework.
	\section{Acknowledgments}
	I would like to thank the M. Shokri because of fruitful discussions and comments. I also appreciate the U. Heinz to encourage me to do this work.
	\appendix
	\section{}
	I want to show that negativeness of transports do not contradict with the second law of thermodynamics. In order to do this,  the following definition for canonical entropy current of charged fluid is used
	\begin{align}\label{eqapp1}
		T S^{\mu}_{can} = p_{id} u^\mu - T^{\mu \nu}u_{\nu} - \mu J^{\mu}.
	\end{align}
	It can be shown that this form of entropy current is invariant under the frame redefinition \cite{Kovtun:2012rj}.  Also the index "id" refers to ideal part of pressure. By putting the corresponding expressions for energy momentum tensor and vector current into the latter relation, after a bit calculation I reach to the following result for canonical entropy
	\begin{align}\label{eqapp2}
		T S^{\mu}_{can} = \left( T s_{id} + \mathcal{E}_{res} - \mu \mathcal{N}_{res}\right) u^{\mu} + \mathcal{Q}^{\mu} - \mu \mathcal{J}^{\mu}.
	\end{align}  
	In the latter relation the $\mathcal{E}_{res}$ and $\mathcal{N}_{res}$ are the resistive parts of energy and number density seen in the relation \eqref{eqsec114} and \eqref{eqsec115}. Definition of $s_{id}$ is as $T s_{id} = \epsilon_0 + p_{id} - \mu n_{0}$ and $\mathcal{Q}^{\mu}$ and $\mathcal{J}^{\mu}$ have no ideal part and start from first order gradient terms according to the relation \eqref{eqsec115}. By using the Equations of Motion (EoM) as in the relations \eqref{eqsec11} and \eqref{eqsec12}, divergence of canonical entropy can be written as
	\begin{align}\label{eqapp3}
		\partial_{\mu} S^{\mu}_{can} = - T^{\mu \nu}_{res} \partial_{\nu} \left( {u_\nu\over T}\right) - \mathcal{J}^{\mu} \partial_{\mu} \left(\mu \over T\right).
	\end{align}  
	The resistive parts of energy momentum tensor and first order gradient terms of current has to be inserted into the latter relation. In these terms the transport coefficients appear along with their bases. For charged conformal matter thanks to the relation  \eqref{eqsec213}, after a little computation  the following off-shell relation is derived 
	\begin{align}\label{eqapp4}
		\partial_{\mu} S^{\mu}_{can}= S^{T} \cdot M_s \cdot S + V^{T}_{\mu} \cdot M_v \cdot V^\mu.
	\end{align}
	The $"T"$ stands for transpose and $S$ and $V$ are scalar and vector bases as it follows
	\begin{align}\label{eqapp5}
		&S= \left({u^\mu \partial_{\mu} T\over T}, \partial_{\mu} u^\mu, u^\mu \partial_{\mu} \left(\mu \over T\right)\right)^{T},\\\label{eqapp6}
		& V^\mu = \left(u^\nu \partial_{\nu} u^\mu, {\Delta^{\mu \nu} \partial_{\nu} T\over T}, \Delta^{\mu \nu} \partial_{\nu} \left(\mu \over T\right)\right)^T.
	\end{align}
	$M_s$ and $M_v$ are also scalar and vector matrices involve the transports
	\begin{align}\label{eqapp7}
	&	M_s = \left(\begin{array}{ccc}
		-3 \epsilon_{1}& -\epsilon_{1} & - {3\nu_1 + \epsilon_{2} \over 2}\\
		-\epsilon_{1}& - {\epsilon_{1}\over 3} &- {\nu_1 \over 2 } - {\epsilon_{2}\over 6}\\
		- {3\nu_1 + \epsilon_{2} \over 2}& - {\nu_1 \over 2 } - {\epsilon_{2}\over 6}& - \nu_2
		\end{array}		
		\right),\\\label{eqapp8}
	&		M_v = \left(\begin{array}{ccc}
		- \theta_{1}& -\theta_{1} & - {\gamma_1 + \theta_{2} \over 2}\\
		-\theta_{1}& - \theta_{1} &- {\gamma_1 + \theta_{2} \over 2}\\
		- {\gamma_1 + \theta_{2} \over 2}& - {\gamma_1 + \theta_{2} \over 2}& - \gamma_2
		\end{array}		
		\right).
	\end{align}
	The transports on this level (the off-shell level) have not to be limited, since transports are frame dependent quantities while the entropy current is a frame independent one. Therefore, the relation \eqref{eqapp4} is studied in the on-shell limit which means that EoM are used to eliminate the dependent vector and scalar bases. These EoM  involve only the ideal part of thermo fields since EoM are first order in gradient and putting the first order terms of constitutive relations, make them to be second order which is beyond the scope of our calculation. Therefore, I use of the following scalar equations for charged fluid by taking the equilibrium values for thermo fields
	\begin{align}\label{eqapp9}
		& u^\mu \partial_{\mu} \epsilon_0 + \left(\epsilon_0 + p_{id}\right) \partial_{\mu} u^\mu =0,\\\label{eqapp10}
		& u^\mu \partial_{\mu} n_{0} + n_{0} \partial_{\mu} u^\mu =0,
	\end{align} 
	to vanish two of scalar bases in favor of another one. I take  the EoS of conformal charged matter in four dimension as
	\begin{align}\label{eqapp11}
		&{\epsilon_0\over T^4} = {3 p_{id}\over T^4} = a + b \left(\mu\over T\right)^2 + c \left(\mu \over T\right)^4,\\\label{eqapp12}
		& {n_{0} \over T^3} = {1\over 3 T^3}{\partial p_{id} \over  \partial \mu} = {2\over 3} \left(2 c \left(\mu \over T\right)^3 +  b \left(\mu \over T\right)\right),\\\label{eqapp13}
		& {s_{id} \over T^3} = {1\over 3 T^3}{\partial p_{id} \over  \partial T} = {2\over 3} \left(2 a  + b \left(\mu \over T\right)^2\right).
	\end{align}   
	The latter relations have to be inserted into the relations \eqref{eqapp9} and \eqref{eqapp10} and the following equations are obtained
	\begin{align}\label{eqapp14}
	& 	\epsilon_0 {u^\mu \partial_{\mu} T\over T} + {3 n_{0} T\over 4} u^\mu \partial_{\mu} \left(\mu \over T\right)= - {\partial_\mu u^\mu\over 3} \epsilon_0,\\\label{eqapp15}
	& n_{0} {u^\mu \partial_{\mu} T\over T} + \chi T u^\mu \partial_{\mu} \left(\mu \over T\right)= - \partial_\mu u^\mu n_{0}.
	\end{align}
	Solutions of equations \eqref{eqapp14} and \eqref{eqapp15} are very simple to derive
	\begin{align}\label{eqapp16}
	 {u^\mu \partial_{\mu} T\over T} =  - {\partial_\mu u^\mu\over 3}	, \qquad u^\mu \partial_{\mu} \left(\mu \over T\right) = 0.
	\end{align} 
	The same work can be done for vector bases by using the following equation
	\begin{align}\label{eqapp17}
		u^\nu \partial_{\nu} u^\mu = - {\Delta^{\mu \nu} \partial_{\nu} p_{id} \over \epsilon_0 + p_{id}}.
	\end{align} 
	The EoS as in the relations \eqref{eqapp11} to \eqref{eqapp13} has to be plugged into the latter relation and finally we arrive to the following equation
	\begin{align}\label{eqapp18}
		- u^\nu \partial_{\nu} u^\mu = {\Delta^{\mu \nu} \partial_{\nu} T\over T} + {3 n_{0} T\over 4 \epsilon_0}  \Delta^{\mu \nu} \partial_{\nu} \left(\mu \over T\right).
	\end{align}
	In the latter relation, the $\Delta^{\mu \nu} \partial_{\nu} \left(\mu \over T\right)$ is eliminated in favor another two. Now,  the equation \eqref{eqapp16} together with relation \eqref{eqapp18} are used to make on-shell the second law of thermodynamics, the relation \eqref{eqapp4}. Calculations show that it reduces to the following result
	\begin{align}\label{eqapp19}
		\partial_{\mu} S^{\mu}_{can} = \left(u^\nu \partial_{\nu} u^\mu + {\Delta^{\mu \nu} \partial_{\nu} T\over T}\right)^2 \, \left(- \left({4 \epsilon_0 \over 3 n_{0} T}\right)^2 \gamma_{2} + {4 \epsilon_0 \over 3 n_{0} T} \left(\gamma_{1} + \theta_{2}\right) - \theta_1\right).
	\end{align}
	As it is evident, the scalar sections do not enter into the second law and only the vector  transports can be limited by using the second law. Therefore,  we can not constrain the scalar transports in this way. This is a weird result and it is because of the conformal symmetry. On the other hand, in order to satisfy the second law we have to have 
	\begin{align}\label{eqapp20}
		- \left({4 \epsilon_0 \over 3 n_{0} T}\right)^2 \gamma_{2} + {4 \epsilon_0 \over 3 n_{0} T} \left(\gamma_{1} + \theta_{2}\right) - \theta_1 \geq 0.
	\end{align}
	We can not say any thing about the individual vector transports, but the combination of them is limited.  It depends on the chosen matter. For example in the high density medium ($\mu \gg T$), it can be shown that $ {4 \epsilon_0 \over 3 n_{0} T} \to \infty$ and therefore $\gamma_{2} \leq 0$. Also in the high temperature limit the similar event happens. Thus, in the extreme limits according to the second law we have definitely a negative transport $\gamma_{2}$, coefficient of  $\Delta^{\mu \nu} \partial_{\nu} \left(\mu \over T\right)$.
	

\begin{thebibliography}{99}
		\bibitem{Romatschke:2017ejr} 
		P.~Romatschke and U.~Romatschke, ``Relativistic Fluid Dynamics In and Out of Equilibrium,''	arXiv:1712.05815 [nucl-th].
		\bibitem{Florkowski:2017olj} 
		W.~Florkowski, M.~P.~Heller and M.~Spalinski, ``New theories of relativistic hydrodynamics in the LHC era,''	Rept.\ Prog.\ Phys.\  {\bf 81}, no. 4, 046001 (2018).
		\bibitem{Jeon:2015dfa} 
		S.~Jeon and U.~Heinz, ``Introduction to Hydrodynamics,''
		Int.\ J.\ Mod.\ Phys.\ E {\bf 24}, no. 10, 1530010 (2015).
		\bibitem{Kovtun:2012rj} 
		P.~Kovtun, ``Lectures on hydrodynamic fluctuations in relativistic theories,'' J.\ Phys.\ A {\bf 45}, 473001 (2012).
		\bibitem{Aad:2012gla} 
		G.~Aad {\it et al.} [ATLAS Collaboration],``Observation of Associated Near-Side and Away-Side Long-Range Correlations in $\sqrt{s_{NN}}$=5.02  TeV Proton-Lead Collisions with the ATLAS Detector,''	Phys.\ Rev.\ Lett.\  {\bf 110}, no. 18, 182302 (2013).
		\bibitem{CMS:2012qk} 
		S.~Chatrchyan {\it et al.} [CMS Collaboration], ``Observation of Long-Range Near-Side Angular Correlations in Proton-Lead Collisions at the LHC,'' Phys.\ Lett.\ B {\bf 718}, 795 (2013).
		\bibitem{Abelev:2012ola} 
		B.~Abelev {\it et al.} [ALICE Collaboration], ``Long-range angular correlations on the near and away side in $p$-Pb collisions at $\sqrt{s_{NN}}=5.02$ TeV,'' Phys.\ Lett.\ B {\bf 719}, 29 (2013).
		\bibitem{Khachatryan:2016txc} 
		V.~Khachatryan {\it et al.} [CMS Collaboration], ``Evidence for collectivity in pp collisions at the LHC,''	Phys.\ Lett.\ B {\bf 765}, 193 (2017).
		\bibitem{Heller:2020anv} 
		M.~P.~Heller, R.~Jefferson, M.~Spaliński and V.~Svensson, ``Hydrodynamic attractors in phase space,''	arXiv:2003.07368 [hep-th].
		\bibitem{Heller:2015dha} 
		M.~P.~Heller and M.~Spalinski, ``Hydrodynamics Beyond the Gradient Expansion: Resurgence and Resummation,'' Phys.\ Rev.\ Lett.\  {\bf 115}, no. 7, 072501 (2015).
		\bibitem{Heller:2018qvh} 
		M.~P.~Heller and V.~Svensson, ``How does relativistic kinetic theory remember about initial conditions?,'' Phys.\ Rev.\ D {\bf 98}, no. 5, 054016 (2018).
		\bibitem{Heller:2016rtz} 
		M.~P.~Heller, A.~Kurkela, M.~Spaliński and V.~Svensson, ``Hydrodynamization in kinetic theory: Transient modes and the gradient expansion,'' Phys.\ Rev.\ D {\bf 97}, no. 9, 091503 (2018).
		\bibitem{Heller:2013fn} 
		M.~P.~Heller, R.~A.~Janik and P.~Witaszczyk, ``Hydrodynamic Gradient Expansion in Gauge Theory Plasmas,'' Phys.\ Rev.\ Lett.\  {\bf 110}, no. 21, 211602 (2013).
		\bibitem{Shokri:2020cxa} 
		M.~Shokri and F.~Taghinavaz, ``Bjorken flow in the general frame and its attractor,''	arXiv:2002.04719 [hep-th].
		\bibitem{McNelis:2020jrn} 
		M.~McNelis and U.~Heinz, ``Hydrodynamic generators in relativistic kinetic theory,'' arXiv:2001.09125 [nucl-th].
		\bibitem{Denicol:2018pak} 
		G.~S.~Denicol and J.~Noronha, ``Hydrodynamic attractor and the fate of perturbative expansions in Gubser flow,'' Phys.\ Rev.\ D {\bf 99}, no. 11, 116004 (2019).
		\bibitem{Blaizot:2017ucy} 
		J.~P.~Blaizot and L.~Yan, ``Fluid dynamics of out of equilibrium boost invariant plasmas,'' 	Phys.\ Lett.\ B {\bf 780}, 283 (2018).
		\bibitem{Strickland:2017kux} 
		M.~Strickland, J.~Noronha and G.~Denicol, ``Anisotropic nonequilibrium hydrodynamic attractor,'' Phys.\ Rev.\ D {\bf 97}, no. 3, 036020 (2018).
		\bibitem{Aniceto:2015mto} 
		I.~Aniceto and M.~Spaliński, ``Resurgence in Extended Hydrodynamics,''
		Phys.\ Rev.\ D {\bf 93}, no. 8, 085008 (2016).
		\bibitem{Basar:2015ava} 
		G.~Basar and G.~V.~Dunne, ``Hydrodynamics, resurgence, and transasymptotics,'' Phys.\ Rev.\ D {\bf 92}, no. 12, 125011 (2015).
		\bibitem{Hiscock:1985zz} 
		W.~A.~Hiscock and L.~Lindblom, ``Generic instabilities in first-order dissipative relativistic fluid theories,''	Phys.\ Rev.\ D {\bf 31}, 725 (1985).
		\bibitem{Hiscock:1987zz} 
		W.~A.~Hiscock and L.~Lindblom, ``Linear plane waves in dissipative relativistic fluids,''	Phys.\ Rev.\ D {\bf 35}, 3723 (1987).
		\bibitem{Israel:1976tn} 
		W.~Israel, ``Nonstationary irreversible thermodynamics: A Causal relativistic theory,''	Annals Phys.\  {\bf 100}, 310 (1976).
		\bibitem{Hiscock:1983zz} 
		W.~A.~Hiscock and L.~Lindblom, ``Stability and causality in dissipative relativistic fluids,'' 	Annals Phys.\  {\bf 151}, 466 (1983).
		\bibitem{Pu:2009fj} 
		S.~Pu, T.~Koide and D.~H.~Rischke, ``Does stability of relativistic dissipative fluid dynamics imply causality?,'' 	Phys.\ Rev.\ D {\bf 81}, 114039 (2010).
		\bibitem{Denicol:2008ha} 
		G.~S.~Denicol, T.~Kodama, T.~Koide and P.~Mota, ``Stability and Causality in relativistic dissipative hydrodynamics,'' J.\ Phys.\ G {\bf 35}, 115102 (2008).
		\bibitem{Kovtun:2019hdm} 
		P.~Kovtun, ``First-order relativistic hydrodynamics is stable,''
		JHEP {\bf 1910}, 034 (2019).
		\bibitem{Bemfica:2019knx} 
		F.~S.~Bemfica, M.~M.~Disconzi and J.~Noronha, ``Nonlinear Causality of General First-Order Relativistic Viscous Hydrodynamics,''
		Phys.\ Rev.\ D {\bf 100}, no. 10, 104020 (2019).
		\bibitem{Bemfica:2017wps} 
		F.~S.~Bemfica, M.~M.~Disconzi and J.~Noronha, ``Causality and existence of solutions of relativistic viscous fluid dynamics with gravity,'' 	Phys.\ Rev.\ D {\bf 98}, no. 10, 104064 (2018).
		\bibitem{Jensen:2012kj} 
		K.~Jensen, R.~Loganayagam and A.~Yarom, ``Thermodynamics, gravitational anomalies and cones,'' 	JHEP {\bf 1302}, 088 (2013).
		\bibitem{Jensen:2012jh} 
		K.~Jensen, M.~Kaminski, P.~Kovtun, R.~Meyer, A.~Ritz and A.~Yarom,
		``Towards hydrodynamics without an entropy current,'' Phys.\ Rev.\ Lett.\  {\bf 109}, 101601 (2012).
		\bibitem{Abbasi:2017tea} 
		N.~Abbasi, F.~Taghinavaz and K.~Naderi, ``Hydrodynamic Excitations from Chiral Kinetic Theory and the Hydrodynamic Frames,''
		JHEP {\bf 1803}, 191 (2018).
		\bibitem{Gradshteyn:2007}
		I.~ S.~ Gradshteyn and I.~M.~Ryznik, ``Table of integrals, series and products,'' Elsevier/Academic Press, Amsterdam, 2007, seventh ed.
	\end{thebibliography}
\end{document}